\documentclass[11pt]{article}
\usepackage{cite}
\usepackage[dvipsnames]{xcolor}
\usepackage[export]{adjustbox}
\usepackage{xcolor}
\usepackage{color,soul}
\usepackage{amsmath,amsfonts,amssymb}
\usepackage{rotate,multicol,wrapfig}
\usepackage{color}
\usepackage{float}
\usepackage{epsfig,epsf}
\usepackage{bm}
\usepackage{dcolumn,subfig}% Align table columns on decimal point
\usepackage{array}
\usepackage{relsize}
\usepackage[bookmarks, breaklinks, colorlinks,urlcolor=blue, citecolor=red, 
linkcolor=blue]{hyperref}
\usepackage[font=small,labelfont=bf]{caption} 
  \usepackage{color}
\usepackage{multirow}
 \usepackage{caption}
 \usepackage{float}
 \usepackage{relsize}
 \usepackage{cancel}
 \usepackage{makecell}
 \usepackage{placeins}
 \usepackage{graphicx}
 \usepackage[utf8x]{inputenc}

 \def\Babar{{\mbox{\slshape B\kern-0.1em{\smaller A}\kern-0.1em B\kern-0.1em{\smaller A\kern-0.2em R}}}}
 \DeclareUnicodeCharacter{2212}{-}
\begin{document}

\renewcommand*{\thefootnote}{\fnsymbol{footnote}}

%%%%%%%%%%%%%%%%%%%%%%%%%%%%%%%%%%%%%%%%%%%%%%%%%%%%%%%%%%%%%%%%%%%%%%%%%%%%%%%%%%%%%%%%%
\begin{center}
 {\Large\bf{ Investigation of $(g-2)_{\mu}$ anomaly  in the $\mu$-specific 2HDM with Vector like leptons and the phenomenological implications}}\\[5mm]
 Md. Raju$^{a,}$\footnote{mdrajuphys18@klyuniv.ac.in}, 
 Abhi Mukherjee$^{a,}$\footnote{abhiphys18@klyuniv.ac.in}, and
 Jyoti Prasad Saha$^{a,}$\footnote{jyotiprasadsaha@gmail.com }\\[3mm]
 {\small\em $^a$Department of Physics, University of Kalyani, Kalyani 741235, India} 
 
 %\today
 \end{center}
%%%%%%%%%%%%%%%%%%%%%%%%%%%%%%%%%%%%%%%%%%%%%%%%%%%%%%%%%%%%%%%%%%%%%%%%%%

%\vspace{5mm}
%%%%%%%%%%%%%%%%%%%%%%%%%%%%%%%%%%%%%%%%%%%%%%%%%%%%%%%%%%%%%%%%%%%%%%%%%%

\begin{abstract}
%%%%%%%%%%%%%%%%%%%%%%%%%%%%%%%%%%%%%%%%%%%%%%%%%%%%%%%%%%%%%%%%%%%%%%%%%%
The anomalous magnetic moment of muons has been a long-standing problem in SM. The current  deviation of experimental value  of the  $(g-2)_{\mu}$ from the standard model prediction is exactly $4.2\sigma$. Two Higgs Doublet  Models can accommodate this discrepancy but such type of model naturally generate flavor changing neutral current(FCNC). To prevent this it was postulated that 2HDM without FCNC  required that all fermions of a given charge  couple to the same Higgs boson but the rule  breaks  in Muon Specific Two Higgs  Doublet Model  where all fermions except muon couple to one Higgs doublet and muon  with the other Higgs doublet. The Muon Specific Two Higgs Doublet model explain muon anomaly with a fine tuning problem of very large $\tan\beta$ value with other parameters. We have found a simple solution of this fine tuning problem by  extending this model with a vector like lepton generation which could explain the muon anomaly at low $\tan\beta$ value with a heavy pseudo scalar Higgs boson under the shadow of current experimental and theoretical constraints. Moreover, with the help of the cut based analysis and multivariate analysis methods, we have also attempted to shed some light on the potential experimental signature of vector lepton decay to the heavy Higgs boson in the LHC experiment. We have showed that a multivariate analysis can increase the vector like leptons signal significance by up to an order of magnitude than that of a cut based analysis.  
\end{abstract}

\setcounter{footnote}{0}
\renewcommand*{\thefootnote}{\arabic{footnote}}
%%%%%%%%%%%%%%%%%%%%%%%%%%%%%%%%%%%%%%%%%%%%%%%%%%%%%%%%%%%%%%%%%%%%%%%%%%%%%%%%%%%%%%%%%%%%%%%%%%%%%%%%%%%%%%%%%%%%%%%%%%%%%%%%%%%%%%%%%%%%%%%%%%%%
\section{Introduction}
The Standard Model (SM)  contribute an amazing interpretation  of nature persisting draconian test at both the current energy and precision frontiers. Due to lack  of any direct signal for new particles at the LHC puts stringent  bounds for different new particles up to several TeV. The remarkable consistency between the predictions from the standard model (SM) and the experimental data from the LHC so far has indicated that the SM is the appropriate effective theory of electroweak (EW) symmetry breaking. The measurement of the magnetic moment of the muon deviates from the SM prediction by more than three standard deviations. The  $(g-2)_{\mu}$ Collaboration of Fermilab  recently published a new result from Run 1 experiment measuring the anomalous magnetic moment of the muon  \cite{Muong-2:2021ojo}. Before this result  the discrepancy between the experimental measurement $a_{\mu}^{exp}$\cite{Muong-2:2006rrc} and the Standard Model $a_{\mu}^{SM}$ prediction was \\
\begin{equation}
\Delta a_{\mu}^{exp} = a_{\mu}^{exp} - a_{\mu}^{SM} =(279 \pm 76) \times 10^{-11} \hspace{2em} (3.7 \sigma)
\end{equation}
while the new combined result is \cite{Muong-2:2021ojo} 
\begin{equation}
\Delta a_{\mu}^{exp}  =(251 \pm 59) \times 10^{-11} \hspace{2em} (4.2 \sigma)
\end{equation}

From long time  people tried very hard to explain the  $(g-2)_{\mu}$ and there are many papers with different models such as supersymmetric models \cite{Joglekar:2013zya,Kyae:2013hda}, left-right symmetric models\cite{PhysRevLett.44.912}, scotogenic models\cite{Ma:2006km}, 331 models\cite{Long:1995ctv}, $L_{\mu} - L_{\tau}$ models\cite{Heeck:2010pg}, seesaw models\cite{Schechter:1980gr}, the Zee-Babu model \cite{Zee:1985rj,Babu:1988ki} whose detail discussions  can be found in \cite{Lindner:2016bgg}. The expansion of SM lepton sector  with vector leptons, is of particular interest \cite{Dermisek:2013gta,Falkowski:2013jya} can explain the discrepancy.In this type of SM extension with vector like leptons(VLLs), muon mixing with the VLLs is required to explain $(g-2)_{\mu}$, and this mixing will change the coupling of Higgs with muon, which will affect not only the Higgs dimuon decay branching ratio, but also the Higgs diphoton decay, which is strongly disfavored by recent collider Higgs data. 
The $(g-2)_{\mu}$ can also be explained using the minimal SM scalar extension Two Higgs Doublet Model (2HDM)\cite{Branco:2011iw,Bhattacharyya:2015nca}. A discrete $Z_2$ symmetry can be used to block the flavor-changing neutral current that occurs in 2HDM, leading to the emergence of four different types of 2HDM, Lepton-specific  Type-I, Type-II, Type-X and Type-Y,(flipped) \cite{PhysRevD.15.1958}. Among them only Type-X and Type-Y  variants are effective to explain the  $(g-2)_{\mu}$. The corresponding model has an enhanced coupling of lepton with  new heavy scalar of 2HDM  it can solve the muon anomaly including the usual one-loop,two-loop contribution from the Barr Zee type diagrams \cite{Wang:2018hnw,Broggio:2014mna}. The Type-II model is  severely constrained by flavor physics and direct searches of extra Higgs bosons because both the charged lepton and down type quark coupling with the new heavy scalar are proportional to $\tan\beta$. In Type-II model the $(g - 2)_{\mu}$ required high value of $\tan\beta$ and light pseudo scalar mass which is disallowed by B-physics observables \cite{Chun:2015xfx}. The flavour limitations are weaker in Type-X 2HDM than in Type-II 2HDM because the lepton couplings are boosted while the quark couplings are suppressed. Among the two variants, the Type-X only fit to explain the existing muon anomaly escaping the flavor constraint without  any fermionic extension. But there is a problem with Type-X model that is  to satisfy the low energy data it  requires  very light pseudo Higgs boson  and large $\tan\beta$ \cite{Wang:2018hnw,Broggio:2014mna}\cite{Chun:2015xfx,Wang:2014sda,Abe:2015oca,Chun:2015hsa} which is also not allowed by B-physics observables. The Type-X parameter space is also highly constrained by the experimental   muon-specific 2HDM model with
VLL measurement of leptonic tau decay. As a result, the parameter region which can explain the discrepancy in the  $(g-2)_{\mu}$ at the $1\sigma$ level is excluded by the constraint from the tau decay\cite{Abe:2015oca}.\\

Due to the shortcomings of Type-II and Type-X models to explain the  $(g-2)_{\mu}$ a new type of 2HDM was proposed by ASY (Tomohiro Abe,Ryosuke Sato and Kei Yagyu)  to probe the  $(g-2)_\mu$ problem \cite{Abe:2017jqo}. The model was structured in such a way that without losing the advantage of type-X model it could accommodate the solution of $(g-2)_{\mu}$. By the implementation of $Z_4$ symmetry they managed to stop the flavor changing neutral current process and simultaneously constrained the model in such a way that  the only second generation of SM leptons couple with one doublet and all others quarks and leptons couple to other doublet. The details of couplings and quantum numbers can be found in \cite{Abe:2017jqo}. The  muon specific 2HDM ($\mu$2HDM) is also important from another point of view as CMS and ATLAS have performed \cite{CMS:2018nak,ATLAS:2019ain} a search for the dimuon decay; the most recent study by ATLAS \cite{ATLAS:2019ain} finds a branching ratio of $0.5\pm 0.7$ times the Standard Model branching ratio (the uncertainty is one standard deviation).  This value is consistent with SM value but if the dimuon decay is not discovered in near future then it will be certain that the branching ratio is substantially below of SM. As the ditau decay is expected to be like SM which means  that the muon and tau must couple with different Higgs doublet\cite{Ferreira:2020ukv}.\\

Above all, these positive aspects, the $\mu$2HDM has a drawback,  it requires very high $\tan\beta$ typically of $O(1000)$ to explain muon anomaly. As usual such large $\tan\beta $ value causes problem with perturbation theory, unitarity, electroweak precision observables, etc. Though ASY shows that this problem can be bypassed if one chooses the free parameters carefully. So in this model the  $(g-2)_{\mu}$ solution is fine-tuned. Though $\mu$2HDM is more acceptable  than the Type-II and Type-X models to explain $(g-2)_{\mu}$, there is  fine tuning problem with very large $\tan\beta$. We have found out a possible solution to this fine tuning problem by adding a vector lepton doublet and singlet with the model $\mu$2HDM and we called it $\mu$2HDM+VLLs. In this model we have applied the ASY mechanism to extended lepton sector and assigned the respective quantum numbers to  vector like leptons . Here the new leptons  couple to $\Phi_{1}$, so only muon can mix with the vector like  lepton. Though in  \cite{Dermisek:2020cod} they have studied the muon anomaly including muon mixing with vector like lepton  but they have  considered the 2HDM Type-II model with the assumption of muon mixing. We have shown that depending on the ASY mechanism the vector like  leptons can spontaneously mix with muon only, which is a unique situation and also it does not carry 2HDM Type-II enhanced quark coupling limitation. Furthermore the charged VLL and charged Higgs will contribute in $ h \rightarrow \gamma \gamma $ decay so we have also shown the allowed parameter space by $ h \rightarrow \gamma \gamma $ within the experimental limit.\\

In this work we have also discussed the signal of all possible VLLs detection channels at LHC. As we know Leptons lighter than 100 Gev are excluded in the earlier search at LEP experiment\cite{L3:2001xsz} and the LHC, ATLAS rejected VLLs that underwent singlet transformation under $SU(2)_L$ in the energy range of 114–176 GeV at $95\%$ CL\cite{ATLAS:2015qoy}. Recent CMS experiment with luminosity 77.4 fb$^{-1}$ at 13 TeV, look for doublet VLLs coupling to third-generation leptons only, discarded all VLLs with masses between 120 and 790 GeV at 95$\%$ CL\cite{CMS:2019hsm}. However, those VLL limits were determined using simplified models. In this work, we investigate the collider signature of extended $\mu$2HDM  with vector like leptons (VLLs), which includes both $SU(2)$ doublet  and a singlet. We analyze this $\mu$2HDM+VLLs in its multi-lepton final state using cut-based analysis as well as multivariate analysis (MVA). This two-way search offers us a comparative examination of the model and provides us the optimised cut value of parameters to find a sensitivity that is compatible to understanding the model. \\

This paper is organized as follows: In sec.\eqref{model} we have discussed the extended $\mu$2HDM model coupled with VLL and also the contribution of this model in the  $(g-2)_\mu$. Sec.\eqref{constraints} contain all the experimental and the theoretical constraints that have been used to constrain our model. In the sec.\eqref{result} we have discussed about the parameter space allowed by all the constraints. The collider phenomena of the VLL in the light of recent collider data and MVA analysis has been discussed in the sec.\eqref{collider}. In the final sec.\eqref{conclusion} we conclude all our findings and some relevant formulas have been showcased in the Appendix.
%%%%%%%%%%%%%%%%%%%%%%%%%%%%%%%%%%%%%%%%%%%%%%%%%%%%%%%%%%%%%%%%%%%%%%%%%%%%%%%%%%%%%%%%%%%%%%%%%%%%%%%%%%%%%%%%%%%%%%%%%%%%%%%%%%%%%%%%%%%%%%%%%%
\vspace{-5mm}
\section{The Model} \label{model}

The scalar  sector of the $\mu$2HDM is made of two $SU(2)_L$ doublet scalar fields $\Phi_1$ and $\Phi_2$. To prevent the FCNCs we apply $Z_4$ symmetry under which the fields transform as $\Phi_1 \rightarrow -\Phi_1$ and $\Phi_2 \rightarrow \Phi_2$. We have extended $\mu$2HDM  with vector like leptons (VLLs) including both $SU(2)$ doublet $L_{L,R}$ and singlet representation $E_{L,R}$. The quantum numbers of SM leptons, Higgs doublets and vector-like fields are represented in Table 1.\\
\vspace{-6mm}
\begin{center}
	\begin{tabular}{|l | c | c | c|c|c |c| c|c|c|c|c|c|}
		\hline
&$l^e_{L}$&$l^{\tau}_{L}$  &$e_{R}$ &$\tau_{R}$ &$\Phi_2$ & $l^{\mu}_{L}$ & $\mu_{R}$ & $\Phi_{1}$ & $L_{L}$ & $L_{R}$ & $E_{L}$ & $E_{R}$   \\             
		\hline
		$Z_4$     &   $1$   &  $1$   &$1$ &   $1$   & $1$   & $i$ & $i$ & $-1$ & $i$ & $-i$ & $-i$  &    $i$  \\  
		\hline
	\end{tabular}
\captionof{table}{Quantum numbers of Standard Model leptons, Higgs doublets, and vectorlike leptons under $Z_4$.} 
\end{center}
The Yukawa interaction  for the muon terms following $Z_4$ symmetry under this charge assignment are given by 
\begin{eqnarray}
\mathcal{L}= - y_{\mu}\bar{l}_L \mu_R \Phi_{1} - \lambda_L\bar{L}_L\mu_R \Phi_{1} - \lambda_E \bar{l}_LE_R \Phi_{1}  -\lambda \bar{L}_LE_R \Phi_{1} \nonumber \\- \bar{\lambda}\Phi_{1}^\dagger \bar{E}_L L_R  -   M_L \bar{L}_L L_R -  M_E \bar{E}_L E_R+  h.c
\end{eqnarray}

The lepton and scalar doublets can be written as,
\begin{eqnarray}
l_L = 
\begin{pmatrix}
\nu_{\mu} &\\
\mu_{L}^- 
\end{pmatrix} ,
L_{L,R} = 
\begin{pmatrix}
L_{L,R}^0 &\\
L_{L,R}^-
\end{pmatrix},
\Phi_{1} = 
\begin{pmatrix}
\Phi_{1}^+ &\\
\Phi_{1}^0
\end{pmatrix},
\Phi_{2} = 
\begin{pmatrix}
\Phi_{2}^0 &\\
\Phi_{2}^-
\end{pmatrix},
\end{eqnarray}

As usual in 2HDM, we have,
\begin{eqnarray}
\Phi_{1}^0 = v_1 + \frac {1}{\sqrt{2}}(-h \sin\alpha + H \cos \alpha  ) + \frac{i}{\sqrt{2}} (G \cos \beta - A \sin \beta)  \\
\Phi_{2}^0 = v_2 + \frac {1}{\sqrt{2}}(h \cos\alpha + H \sin \alpha  ) - \frac{i}{\sqrt{2}} (G \sin \beta + A \sin \beta)   \\
\Phi_{1}^{\pm} = \cos\beta G^{\pm} - \sin\beta H^{\pm} ,
\Phi_{2}^{\pm} = - \sin\beta G^{\pm} - \cos\beta H^{\pm}
\end{eqnarray}
The 2HDM possesses five physical Higgs bosons a charged pair ($H^{\pm}$) two neutral $CP$-even  sealars (h and H) and a neutral $CP$-odd scalar (A), often called a pseudoscalar.In paradigm of alignment limit the h is like SM Higgs boson.The charged gauge eigenstate $\phi^{+}_1$ and $\phi^{+}_2$ will give rise to one charged Higgs $H^+$ and a charged Goldstone boson($G^\pm$).The scalar sector of 2HDM is described in detail in \cite{Branco:2011iw}.After  symmetry breaking the vacuum expectation value associated to the neutral components $<\phi^{0}_1>=v_1$ and $<\phi^{0}_2>=v_2$. Here, we have taken into account the following condition $\sqrt{v_1^2 + v_2^2}=v=174$ GeV between the vevs  and we have define $\tan\beta = \frac{v_2}{v_1}$. As a result, the mass matrix is transformed to
	\begin{equation}
	\begin{pmatrix}
		\bar{\mu}_L & \bar{L}^-_L & \bar{E}_L
	\end{pmatrix} M_e \begin{pmatrix}
		\mu_R \\
		L^-_R \\
		E_R
	\end{pmatrix}  = \begin{pmatrix}
		\bar{\mu}_L & \bar{L}^-_L & \bar{E}_L
	\end{pmatrix} \begin{pmatrix}
		y_{\mu} v_1 & 0  &   \lambda_E v_1  \\
		\lambda_L v_1 & M_L    & \lambda v_1 \\
		0  &   \bar{\lambda} v_1  &    M_E
	\end{pmatrix} \begin{pmatrix}
		\mu_R \\
		L^-_R \\
		E_R
	\end{pmatrix} \\
	\end{equation}
	We need to diagonalize the mass matrix using
	\begin{equation}
	U^{e\dagger}_L \begin{pmatrix}
		y_{\mu} v_1 & 0  &   \lambda_E v_1  \\
		\lambda_L v_1 & M_L    & \lambda v_1 \\
		0  &   \bar{\lambda} v_1  &    M_E
	\end{pmatrix}  U^e_R = \begin{pmatrix}
		m_\mu   & 0 & 0  \\
		0  &   m_{e_4}  & 0   \\
		0    & 0   & m_{e_5}   
	\end{pmatrix}  
\end{equation}
We have  two mass eigenstate $e_4$ and $e_5$ of charged vector leptons. Simultaneously one part of the muon mass came from the yukawa term, and the other part cames from mixing with VLLs. For the sake of simplicity, we will use the neutral vector lepton  ($\nu_4$)  mass, which is provided by $M_L$.\\
We can obtain the effective lagrangian from equation (3) in the heavy mass limit of VLLs, which will put some light on the corelation of muon mass from VLLs mixing and the impact on  $(g-2)_{\mu}$.
\begin{equation}
	\mathcal{L} = - y_{\mu} \bar{l}_{L} \mu_R H_1 - \frac{\lambda_L \bar{\lambda} \lambda_E}{M_L M_E}\bar{l}_{L} \mu_R H_1 H_1^{\dagger} H_1 + h.c
\end{equation}
where the second term of this equation quantify a new source of muon mass due to VLLs \cite{Dermisek:2020cod}. By following this route, we may roughly estimate the contributions from each diagram in Figure~(1) to obtain the transparent effect of the new Muon mass source on $(g-2)_{\mu}$ .\\
The contribution from all  1-loop diagrams can be expressed in a single formula as 
\begin{equation}
	\Delta^i_{\mu} \simeq \frac{k^i}{16\pi^2}\frac{m_{\mu} m_{\mu}^{LE}}{v^2} , m_{\mu}^{LE} \equiv \frac{\lambda_L \bar{\lambda} \lambda_E}{M_L M_E}v^3 \cos^3\beta
\end{equation} 
where $k^W = 1$ ,$k^Z = -\frac{1}{2}$ , $k^h = -\frac{3}{2}$,$k^H = -\frac{11}{12} \tan^2\beta$ , $k^A = -\frac{5}{12}\tan^2\beta$ and $k^{H^{\pm}} = \frac{1}{3}\tan^2\beta$ these are good approximation with heavy VLLs and $M_L \simeq M_E \simeq m_{H,A,H^{\pm}}$.
To explain the  $(g-2)_{\mu}$ in $\mu$2HDM requires very high value of $\tan\beta$ typically of O(1000). Usually such large value of $\tan\beta$ causes concern with perturbative theory, unitarity and electroweak precision observables etc. So to circumvent  the large  $\tan\beta$ problem in $\mu$2HDM we have extended the lepton sector with VLLs. Further we have  restricted  the VLLs mixing with muon only. Then from the orginal lagrangian eqn (3) we have built an effective lagrangian (10). So now we have two sources of muon mass, one from the typical yukawa mass term, and the other from the mixing with VLLs, which is a novel source of muon mass. This new mass parameter will linearly modify  the muon yukawa couplings and generate a remarkable effect on  $(g-2)_{\mu}$ correction.
Now if we see the approximate contribution of heavy Higgs in (11) the factor $m_{\mu}^{LE}$ due to VLLs is not present in the $\mu$2HDM and this factor plays an interesting role for heavy Higgs contribution.
%%%%%%%%%%%%%%%%%%%%%%%%%%%%%%%%%%%%%%%%%%%%%%%%%%%%%%%%%%%%%%%%%%%%%%%%%%%%%%%%%%%%%%%%%%%%%%%%%%%%%%%%%%%%%%%%%%%%%%%%%%%%%%%%%%%%%%%%%%%%%%%%%%
\subsection{Contribution in  $(g - 2)_{\mu}$ } \label{muon}
The contribution for  $(g-2)_{\mu}$ anamoly is completely derived in the present model and other variation of 2HDM have presented in \cite{Dermisek:2021ajd}.
The mixing of vector like leptons with muon generate new diagramms  for $(g-2)_{\mu}$ at one loop level which shown in Fig.(1).The one-loop diagramms are dominant compared to two loop Bar-Zee (BZ) diagramms with heavy fermions in the loop.The mixing of muon with new vector like lepton generate an additional contribution to $(g-2)_\mu$.The contribution comming from W and Z bosons \cite{Dermisek:2013gta} are. \\
The W boson contribution is
\begin{eqnarray}
	\Delta a_{\mu}^W=\frac{m_{\mu}}{16 \pi^2 M_W^2} \sum_{a=4,5}	\Big[m_{\mu}\Big(|g_R^{W\nu_{a}\mu}|^2 + |g_L^{W\nu_{a}\mu}|^2 \Big) F_W(x_W^a) \nonumber  \\ -m_{\nu_{a}} Re[g_R^{W\nu_{a}\mu}( g_L^{W\nu_{a}\mu})^*] G_W(x_W^a)\Big]
\end{eqnarray}

The Z-boson contribution to $(g-2)_{\mu}$ is then given by
\begin{eqnarray}
	\Delta_{\mu}^Z=\frac{-m_{\mu}}{8 \pi^2 M_Z^2}\sum_{a=4,5}\Big[m_{\mu}\Big(|g_R^{Z_{\mu e_a}}|^2 + |g_L^{Z_{\mu e_a}}|^2 \Big) F_Z(x_W^a) \nonumber \\
	-m_{e_{a}} Re[g_R^{Z_{\mu e_a}}(g_R^{Z_{\mu e_a}})^*] G_Z(x_Z^a)\Big]
\end{eqnarray}

The contribution from neutral Higgs bosons $h , H$ and $A$ are identical in nature except their  couplings factors. For $\phi = h,H,A$  we can define the couplings of charged leptons to neutral Higgses by 
\begin{eqnarray}
	\Delta_{\mu}^{\phi}=\frac{m_{\mu}}{32 \pi^2 m_{\phi}^2}\sum_{a=4,5}\Big[m_{\mu} \Big(|g^{\phi}_{\mu e_a}|^2 + |g^{\phi}_{ e_a \mu}|^2 \Big ) F_{\phi}(x_{\phi}^a)   \nonumber \\
	+ m_{e_{a}} Re[g^{\phi}_{\mu e_a} g^{\phi}_{ e_a \mu}] G_{\phi}(x_{\phi}^a)\Big]
\end{eqnarray}

The contribution to $(g-2)_{\mu}$ from loops with the charged Higgs is then given by
\begin{eqnarray}
	\Delta_{\mu}^{H^{\pm}}=\frac{- m_{\mu}}{16 \pi^2 m_{H^{\pm}}^2}\sum_{a=4,5}\Big[m_{\mu} \Big(|g^{H^{\pm}}_{\nu_a \mu}|^2 + |g^{H^{\pm}}_{ \mu \nu_a }|^2 \Big ) F_{H^{\pm}}(x_{H^{\pm}}^a)   \nonumber \\
	+ m_{\nu_{a}} Re[g^{H^{\pm}}_{\nu_a \mu} g^{H^{\pm}}_{ \mu \nu_a}] G_{H^{\pm}}(x_{H^{\pm}}^a)\Big]
\end{eqnarray}

\begin{figure}[H]
\centering
{\includegraphics[width=0.75\textwidth]{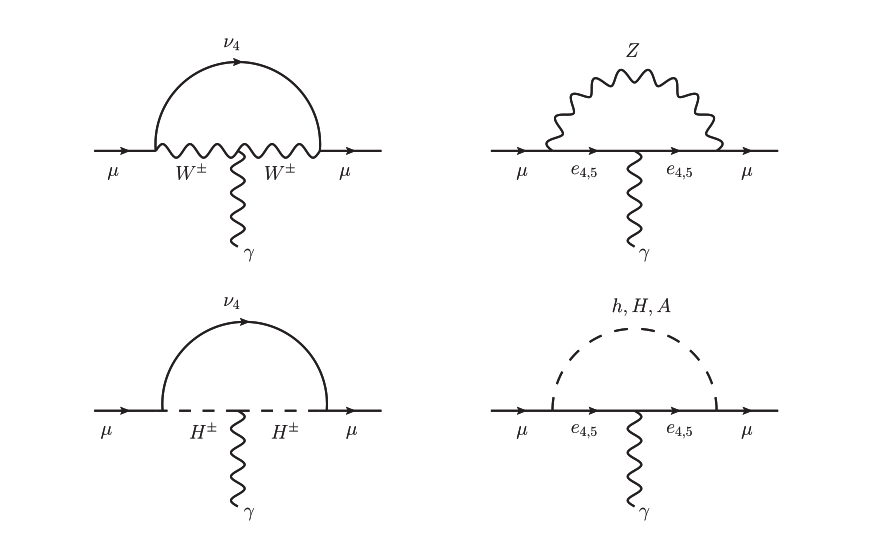}} 
\caption{\emph{The diagrams contributing the muon anomalous magnetic moment with $W, Z, h, H, A, H^{\pm}$}}
\label{fig:foobar}
\end{figure}
%%%%%%%%%%%%%%%%%%%%%%%%%%%%%%%%%%%%%%%%%%%%%%%%%%%%%%%%%%%%%%%%%%%%%%%%%%%%%%%%%%%%%%%%%%%%%%%%%%%%%%%%%%%%%%%%%%%%%%%%%%%%%%%%%

\section{ Constraints on the  Model Parameters} \label{constraints}
\subsection{ Constraints from the Z pole measurements}
\vspace{-5mm}
The $\mu$2HDM+VLLs model allowed the mixing of only  second generation lepton  with the vector like leptons leading to modifications of muon couplings to W and Z bosons, this modification can affect the different observables like $\mu$ lifetime, the forward-backward and left-right asymmetries involving muons,the Z width into $\mu^+ \mu^-$ and $\nu_{\mu} \bar{\nu}_{\mu}$. The EW measurements constrain possible modification of couplings of the muon to the Z and W bosons\cite{ALEPH:2005ab,ParticleDataGroup:2020ssz}. For the $\mu$2HDM+VLLs, the leptons as well as the VLLs couple exclusively to $\Phi_{1}$ so that the global electroweak fit for the vector like leptons which gives the following limit \cite{Kannike:2011ng,Crivellin:2020ebi}.
\begin{eqnarray}
	\Big|\frac{\lambda_E v_1}{M_E}\Big|\lesssim 0.03 \\
	\Big|\frac{\lambda_L v_1}{M_L}\Big|\lesssim 0.04 
\end{eqnarray}
%%%%%%%%%%%%%%%%%%%%%%%%%%%%%%%%%%%%%%%%%%%%%%%%%%%%%%%%%%%%%%%%%%%%%%%%%%%%%%%%%%%%%%%%%%%%%%%%%%%%%%%%%%%%%%%%%%%%%%%%%%%%%%%%%%%%%%%%%%%%%%%%%%%%
\subsection{Higgs Diphoton Decays with Vector Lepton}

In the domain of alignment limit the tree level couplings to leptons and gauge bosons become exactly like SM. Along with the charged scalar $H_{\pm}$ of the 2HDM, the charged VLLs can contribute to the loop-induced decay mode of the Higgs into $\gamma \gamma$. As a result, our model must be compatible with the present Higgs to diphoton decay limit. The Higgs to diphoton decay width is expressed including the contribution coming from new particles (vector like leptons) in the loop as
\begin{equation}
	\Gamma_{h\rightarrow \gamma \gamma} =\frac{G_{F} \alpha^2 m_h^3}{128\sqrt{2}\pi^3}\Bigg| \kappa_V F_1(x_W)+\frac{4}{3}\kappa_{ht\bar{t}} F_{\frac{1}{2}}(x_t) + F_{\frac{1}{2}}(x_l) \kappa_{hl\bar{l}}+  \kappa_{hH^{+} H^{-}} F_{+}(x_{H^{\pm}})\Bigg|^2 	
\end{equation}
where $x_j = (\frac{2m_j} { m_h })^2 , (j = W,t,f, H_{\pm} $), $m_h$ is the SM Higgs mass, $\kappa_{hl\bar l}$ ($\kappa_{hH^+ H^-} $) are the
couplings of SM Higgs boson to vector-like fermions (charged Higgs) with mass $M_L (m_+)$ respectively. 
The corresponding  loop functions $F_1 , F_{1/2}$ and $F_+$ which appear in the calculation are 
\begin{eqnarray}
	F_1(x)=2 + 3 x +3 x (2-x)f(x) \\  \nonumber 
	F_{1/2}(x)=-2x[ 1+ (1-x)f(x)] \\   \nonumber 
	F_{+}(x)=-x[1-x f(x)] \\ \nonumber 
	f(x)=\Bigg\{[\sin^{-1} (1/\sqrt{x})]^2 , \   x \geq 1
\end{eqnarray}
\[
f(x)= 
\begin{cases}
(\sin^{-1} (1/\sqrt{x}))^2,&  x\geq 1\\
-\frac{1}{4}(\ln(\frac{1+\sqrt{1-x}}{1-\sqrt{1-x}})-i \pi)^2,  & x<1
\end{cases}
\]
and, the charged Higgs couplings to the SM Higgs is given by \cite{Djouadi:1996yq}
\begin{equation}
	k_{hH^+ H^-}=-\frac{1}{2m_+^2} \Big[\frac{(m_h^2 - 2m_+^2)\cos(\alpha- 3\beta)+ (3m_h^2 + 2m_+^2 -4 m_0^2)\cos(\alpha + \beta)}{4\sqrt{2}  \sin\beta\cos\beta}\Big]
\end{equation}
where
\begin{equation}
	m_0^2 = \frac{m_{12}^2}{\sin\beta \cos\beta}
\end{equation}
We utilized the existing experimental limit to accomplish this. The present experimental limit on the strength of the Higgs to diphoton signal is quite close to its SM value and stands at $\mu_{\gamma \gamma} = \frac{\mu_{\gamma \gamma}^{exp}}{\mu_{\gamma \gamma}^{SM}}=1.18^{+0.17}_{-0.14} $\cite{CMS:2018piu}.
We may now define the ratio of decay width as  describing the enhacement and supression in $h \rightarrow \gamma \gamma $ channel because the VLLs no longer contributes to Higgs production .

\begin{figure}[h!]
   \hspace{.5cm}\subfloat[]{\includegraphics[width=0.40\textwidth]{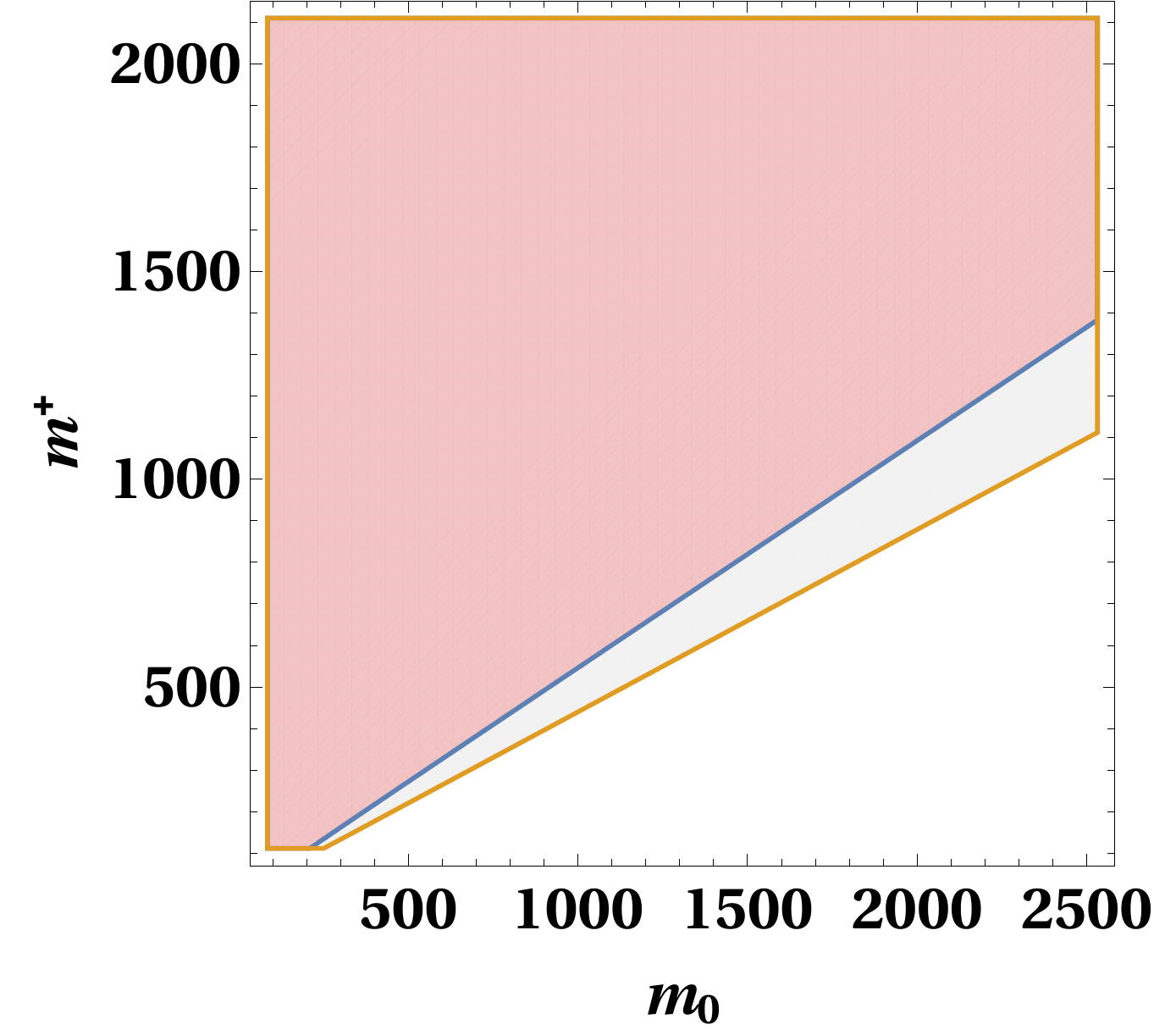}}
	\hspace{3cm}\subfloat[]{\includegraphics[width=0.40\textwidth]{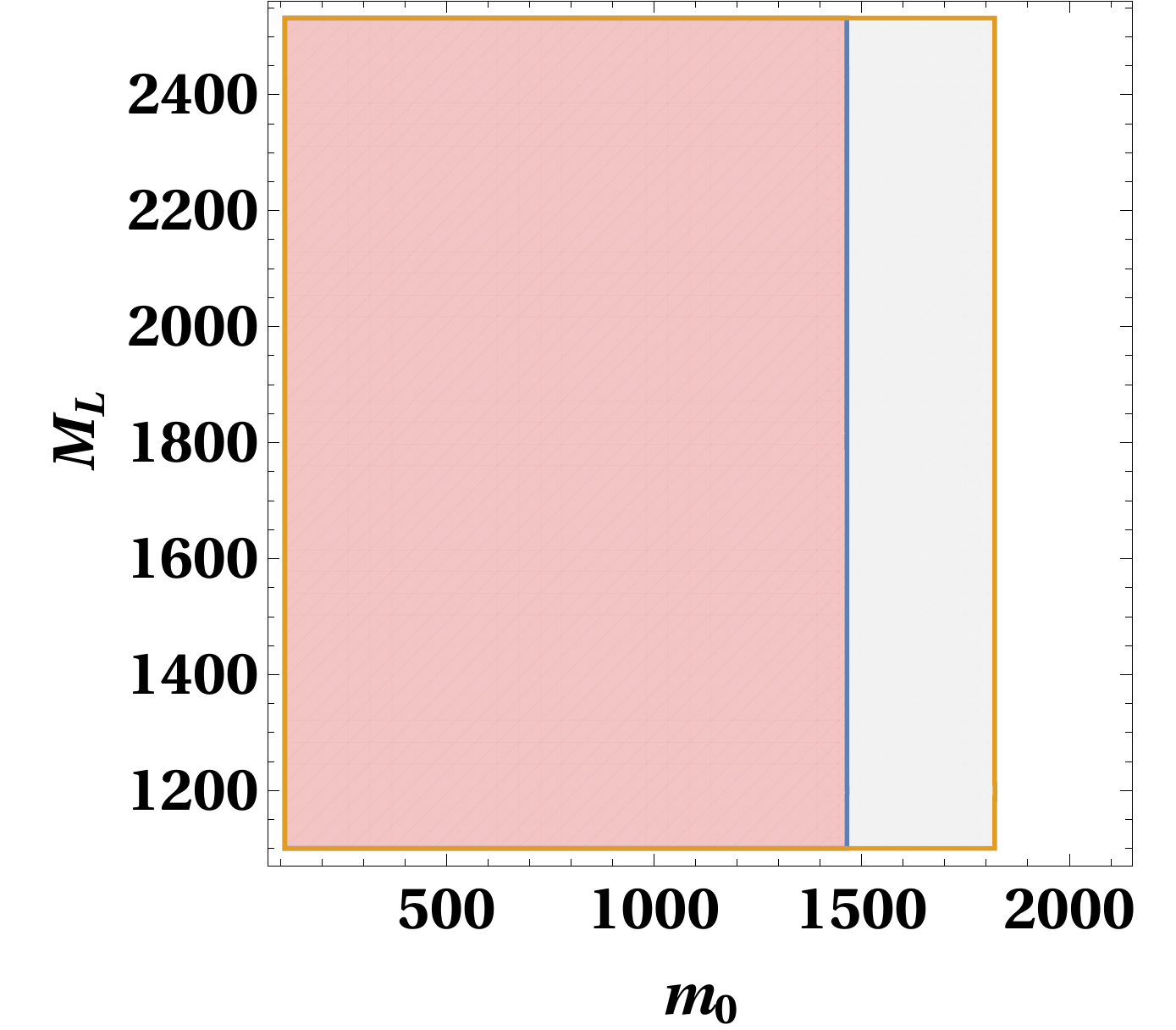}}
	\caption{\emph{Restriction on ($m_0 - m+ $ ) (left) and ($m_0$ - $M_L$ ) (right) . The  LightRed is 1$\sigma$ and Gray is 2$\sigma$ allowed region of  $h\rightarrow \gamma \gamma$ signal strength. Here $m_0$ is the soft breaking parameter defined in the text, and ML is the  mass for vector-like charged lepton.}}
	\label{fig:figure2}
\end{figure}	

%\begin{equation}
%	\mu_{\gamma \gamma}=\frac{\sigma(pp \rightarrow h)}{\sigma_{SM}(pp %\rightarrow h)} \frac{\Gamma(h \rightarrow \gamma \gamma)}{\Gamma_{SM}(h %\rightarrow \gamma \gamma)}=\frac{\Gamma(h \rightarrow \gamma %\gamma)}{\Gamma_{SM}(h \rightarrow \gamma \gamma)}
%\end{equation}
The constraints on mass parameters are demonstrated in Fig.2  for a specific set of parameters (as given in Table 2).These graphs shows that by fine-tuning the soft-symmetry breaking term $m_0$, the charged Higgs mass $m_+$, and the VLL mass parameter $M_L$, the experimental data  can easily be satisfied. It should be emphasised that, while we set a certain value for the VLL Yukawa coupling and $\tan\beta$, the correlation between the mass parameters is unaffected by our choice. 
%%%%%%%%%%%%%%%%%%%%%%%%%%%%%%%%%%%%%%%%%%%%%%%%%%%%%%%%%%%%%%%%%%%%%%%%%%%%%%%%%%%%%%%%%%%%%%%%%%%%%%%%%%%%%%%%%%%%%%%%%%%%%%%%%%%%%%%%%%%%%%%%%%%%

\subsection{Constraints from electroweak precision observables}

Important limitations come from the electroweak oblique parameters, since the extra scalars and leptons contribute to gauge boson masses via loop corrections, in addition to the Higgs data and the theoretical constraints established in the preceding subsections. The scalar contributions to the oblique T and S parameters are well-known, as shown in  \cite{Grimus:2007if,Grimus:2008nb}.The Z and W couplings with VLLs can be written as,
\begin{eqnarray}
	\mathcal{L}^Z = \Big(\bar{f}_{La} \gamma^{\mu} g_L^{Z f_a f_b} f_{Lb} + \bar{f}_{Ra} \gamma^{\mu} g_R^{Z f_a f_b} f_{Rb}\Big)Z_{\mu}  \\
	\mathcal{L}^W = \frac{g}{\sqrt{2}}\Big( \bar{\nu_\mu} \gamma^{\mu} \mu_{L}  +  \bar{L}^0_L \gamma^{\mu} L^-_L + \bar{L}^0_R \gamma^{\mu} L^-_R      \Big)W^+_{\mu} + h.c
\end{eqnarray}
\subsubsection{T-Parameter}

The additional fermion contribution formula \cite{Chen:2017hak}
\begin{eqnarray}
\Delta T_F =\frac{1}{8 \pi s_W^2 c_W^2}\sum_{a,b=2,4,5}\Big[ \big(|g_L^{W\nu_a e_b}|^2 + |g_R^{W\nu_a e_b}|^2\big) \theta_{+}(f_a,f_b) + 2Re\big(g_L^{W\nu_a e_b}  g_R^{W\nu_a e_b*}\big) \theta_{-}(f_a,f_b)   \nonumber \\ -\frac{1}{2} \big(|g_L^{Ze_a e_b}|^2 + |g_R^{Z e_a e_b}|^2\big) \theta_{+}(f_a,f_b) + 2Re\big(g_L^{Z e_a e_b}  g_R^{Z e_a e_b *}\big) \theta_{-}(f_a,f_b) \Big]
\end{eqnarray}
where, $f_a =\frac{m_{e_{a}}^2}{M_z^2}$and the functions are defined as

\[
\theta_{+}(x,y)= 
\begin{cases}
	\frac{x+y}{2}-\frac{x y}{x-y} \ln(\frac{x}{y}),& \text{if } x\neq y \\
	0,              & \text{if } x = y
\end{cases}
\]
\[
\theta_{-}(x,y)= 
\begin{cases}
	\sqrt{x y} \Big[  \frac{x+y}{x -y} \ln(\frac{x}{y}) -2  \Big],& \text{if } x\neq y \\
	0,              & \text{if } x = y
\end{cases}
\]

\subsubsection{S-Parameter}

The general expression for the S-parameter contribution from additional fermions is \cite{Chen:2017hak},
\begin{eqnarray}
\Delta S_F =\frac{1}{2 \pi}\sum_{a,b=2,4,5}\Big[ \big(|g_L^{W\nu_a e_b}|^2 + |g_R^{W\nu_a e_b}|^2\big) \psi_{+}(f_a,f_b) + 2Re\big(g_L^{W\nu_a e_b}  g_R^{W\nu_a e_b*}\big) \psi_{-}(f_a,f_b)   \nonumber \\ -\frac{1}{2} \big(|g_L^{Ze_a e_b}|^2 + |g_R^{Z e_a e_b}|^2\big) \xi_{+}(f_a,f_b) + 2Re\big(g_L^{Z e_a e_b}  g_R^{Z e_a e_b *}\big) \xi_{-}(f_a,f_b) \Big]
\end{eqnarray}
where, $f_a =\frac{m_{e_{a}}^2}{M_z^2}$and the functions are defined as
\begin{eqnarray}
	\psi_{+}(x,y) = \frac{1}{3} -\frac{1}{9}\ln(\frac{x}{y}) \nonumber \\
	\psi_{-}(x,y) = -\frac{x + y}{6 \sqrt{x y}}
\end{eqnarray}

\[
\xi_{+}(x,y)= 
\begin{cases}
	\frac{5 (x^2 + y^2) - 22 x y }{9 (x- y)^2} + \frac{ 3 x y (x + y ) - x^3 - y^3 }{3 (x-y)^3} \ln(\frac{x}{y}),& \text{if } x\neq y \\
	0,              & \text{if } x = y
\end{cases}
\]

\[
\xi_{-}(x,y)= 
\begin{cases}
	\sqrt{x y} \Big[  \frac{x+y}{6 x y}  - \frac{x + y }{(x - y)^2} + \frac{2 x y }{(x - y)^3}\ln(\frac{x}{y})  \Big],& \text{if } x\neq y \\
	0,              & \text{if } x = y
\end{cases}
\]
The current global electroweak fit yields \cite{ParticleDataGroup:2018ovx} 
\begin{equation}
\Delta T	= 0.07 \pm 0.12 , \Delta S = 0.02 \pm 0.07 
\end{equation}
\begin{figure}[H]
\centering
\hspace{0.1cm}\subfloat[]{\includegraphics[width=0.5\textwidth]{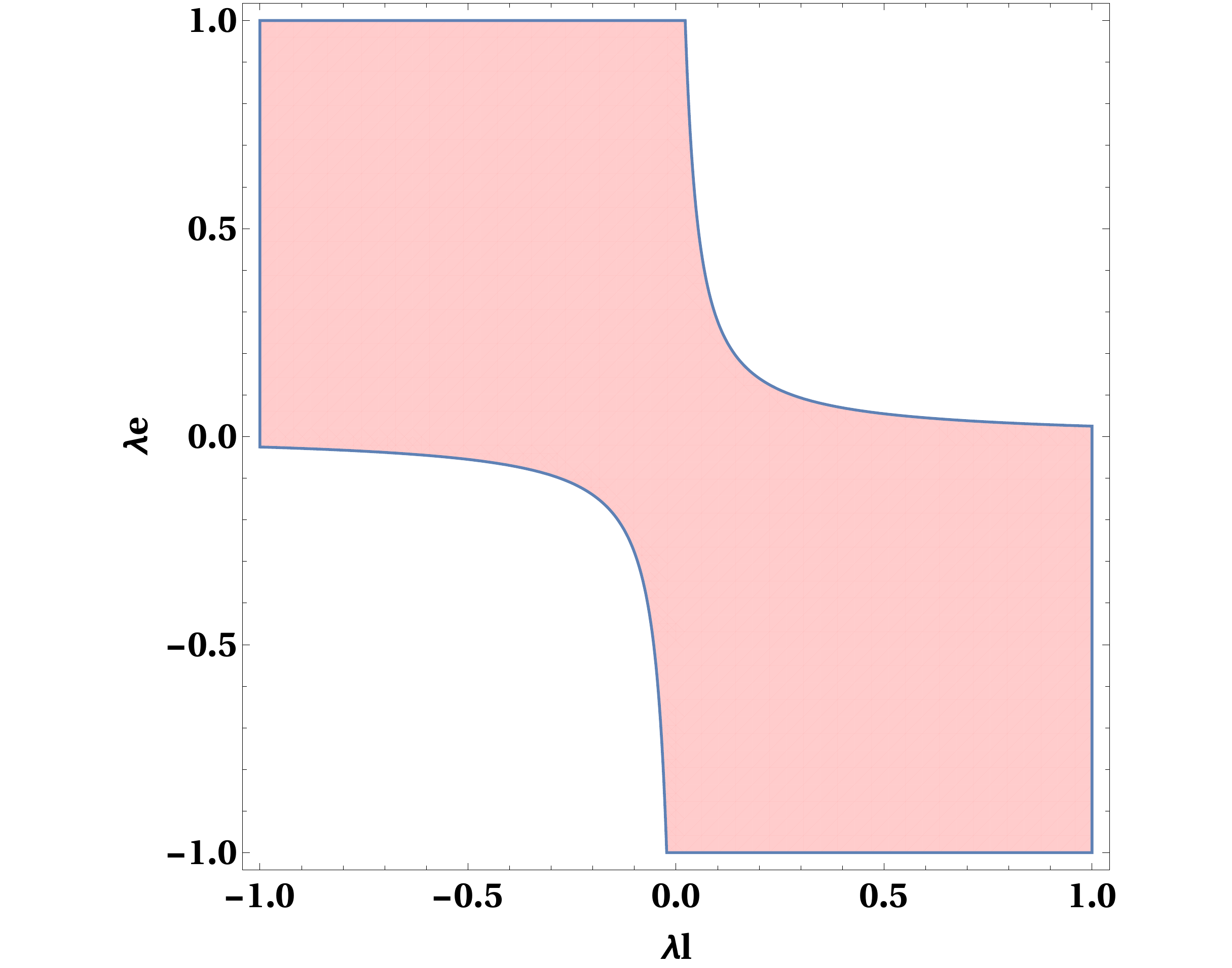}}
\vspace{0.1cm}\subfloat[]{\includegraphics[width=0.5\textwidth]{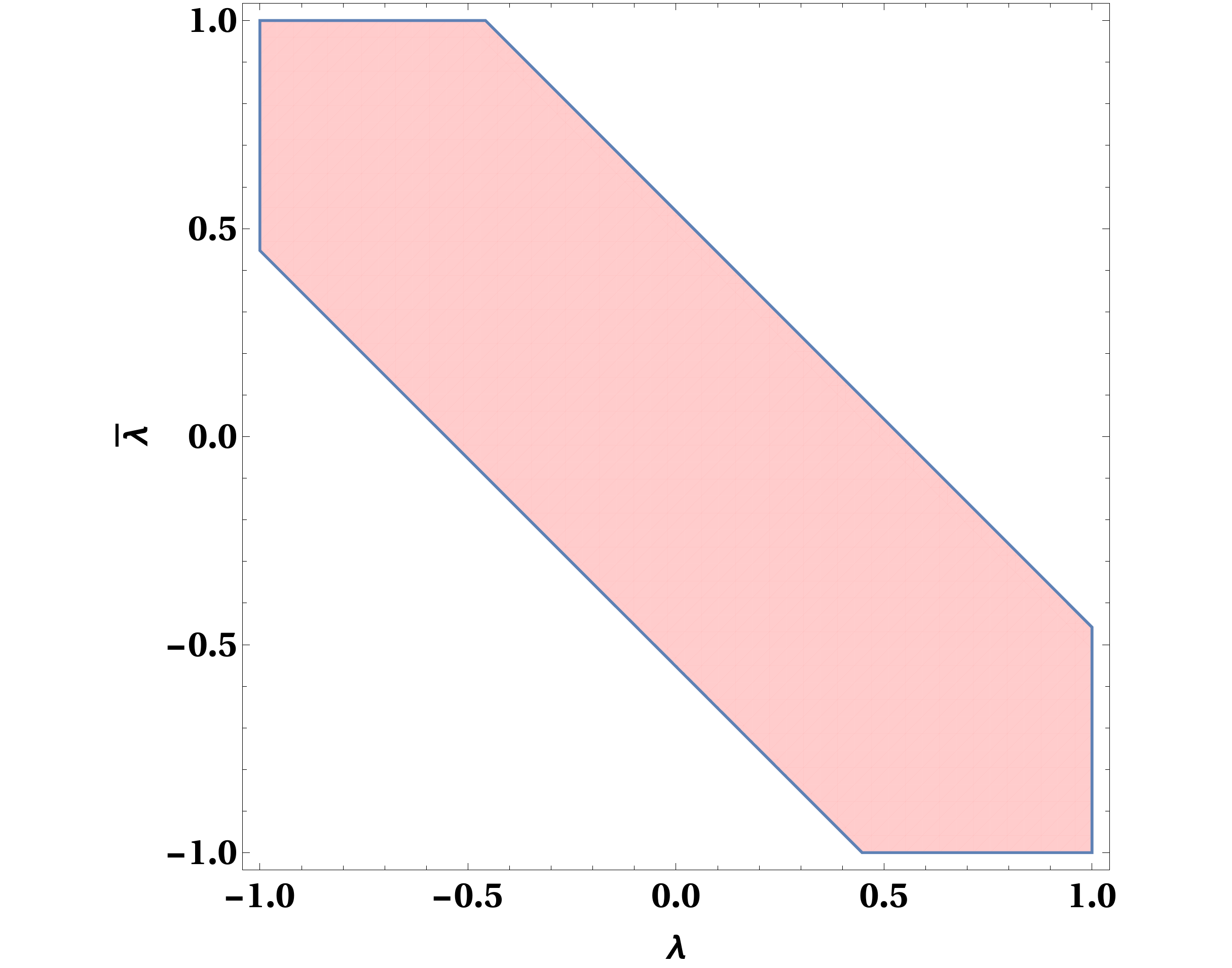}}
\caption{\emph{Allowed region of vector leptons couplings by the S and T  parameters bound .}}
\label{fig:foobar}
\end{figure}
The precision observables can confirm the mixing of SM leptons and VLLs. The coupling $\lambda$ and $\bar{\lambda}$, which mixes the VLLs among themselves, is not constrained by the Z pole observables. Because the mass eigenstates of the heavy charged leptons are dependent on these couplings, these couplings can generate a mass gap between the neutral and charged components of the doublet. This can give correction to oblique T parameter \cite{Peskin:1990zt,Peskin:1991sw} and can be constrained. So we have shown in Fig.\textcolor{red}{3} the allowed parameter space following the oblique S and T parameter constraints. The interesting thing is that for large mass gaps, all values of $\lambda$s are allowed, but when the mass gap between charged and neutral lepton $(M_L -M_E) = \Delta m = $3 GeV) is less than $\Delta m$ the parameters space is disallowed. The scalar also has an impact on oblique parameters that must be considered. Even so, it has been demonstrated that by making the charged Higgs degenerate with the heavy scalar or the pseudo-scalar, the oblique corrections from the scalar sector of 2HDM can be minimised \cite{Grimus:2007if,Grimus:2008nb}. Moreover, the tree level contribution of the charged Higgs exchange diagram to the leptonic decay process is insignificant in the existing $\mu$2HDM model. This is because of the cancellation of the $\tan\beta$ dependency, according to the muon-specific feature of this model.

%%%%%%%%%%%%%%%%%%%%%%%%%%%%%%%%%%%%%%%%%%%%%%%%%%%%%%%%%%%%%%%%%%%%%%%%%%%%%%%%%%%%%%%%%%%%%%%%%%%%%%%%%%%%%%%%%%%%%%%%%%%%%%%%%%%%%%%%%%%%%%%%%%%%%%%%%%%%%%%%%%%%%%%%%%%%%%%%%

\section{Results} \label{result}

The analysis and interpretation of our study are the focus of this section. In order to describe the $(g- 2)_{\mu}$, we will first explore the promising out come and efficiency of the existing $\mu$2HDM  model. The $\mu$2HDM model which is singular  from all existing variant of 2HDM regarding the yukawa structure of leptons.

In $\mu$2HDM model, there is a suppression of both the tau and bottom yukawa couplings. As a result of which the loop factor suppresses the two loop contributions from Bar-Zee type diagrams in $\mu$2HDM model. Therefore, these diagrams are irrelevant in comparison to one-loop contribution. In context of this, we have only considered the one-loop contribution in the following analysis of $(g-2)_\mu$. For the purpose of further discussion and illustrative numerical analysis we have selected some benchmark the parameters (see table\eqref{tableparam}) which satisfy all coupling constraints, Higgs to diphoton data, and the oblique parameter constraints, as mentioned in the past sections. Moreover, one may choose $\sqrt{4\pi}$ as the limit from perturbativity at the scale of new physics to keep the yukawa coupling under perturbative control for large energy scales. However, We have taken into account all yukawa couplings up to the value 1 in our numerical analysis for the same purpose. For the all numerical analysis we have considered $M_{L} > 600 $ GeV and $ M_{E} > 500 $ GeV  to typically satisfy constraints from direct searches for new leptons \cite{CMS:2019hsm,CMS:2018iaf}. Further we have considered $ m_H = m_A = m_{H^{\pm}}$ and applied limits on $H(A) \rightarrow \tau^+ \tau^- $ \cite{ATLAS:2020zms} and $H^+ \rightarrow t \bar{b}$ \cite{ATLAS:2021upq} which are currently the strongest at small and large $\tan\beta$ respectively. These limits are also sufficient to satisfy indirect constraints from flavour physics observables \cite{Haller:2018nnx}. We have also imposed the constraints from $h \rightarrow \mu \mu $ \cite{ATLAS:2020fzp} and muon electroweak data such as Z-pole observables ,the W partial width and the muon life time and constraints from oblique corrections,\cite{ParticleDataGroup:2020ssz}.\\
\begin{table}[h!]
\begin{center}
	\begin{tabular}{ |c|c|c|c|c|c|c|c|c|c|c|c| } 
		\hline
		Parameters & $\tan\beta$  & $m_0$ & $M_H$& $M_A$ & $M_{H^{\pm}}$& $M_L$ & $M_E$ & $\lambda$& $\lambda_e$ & $\lambda_l$& $\lambda_b$   \\ 
		\hline
	Fixed Vlaue & 10 & 800     &  800  & 800  &  800 &  1500 & 1200   &0.0   & 0.5  & 0.5    &-0.5  \\ 
		\hline
	\end{tabular}
\end{center}
\caption{\emph{Fixed parameter values used to generate the allowed region }}
\label{tableparam}
\end{table}

In order to explain the $(g-2)_{\mu} $ we have to take those values of the parameters $ m_{\mu}^{LE} $ (see eqn.11) and $\tan\beta$, so that they satisfy the relation $\tan^2\beta m_{\mu}^{LE} =-m_{\mu}$.
This condition will enhance the contribution that are coming from the H, A and $H^{\pm}$ in the $(g-2)_\mu$ \cite{Dermisek:2020cod}. The enhancement caused by $m_{\mu}^{LE}$ for the low $\tan\beta$ region ($\tan\beta \sim 6-12$) is significantly greater than the improvement for the $\tan^2\beta$ value. For this behavior at low $\tan\beta$ value the heavy Higgs contribute reasonably to accommodate the  $(g-2)_{\mu}$. Also due to the $\tan^2\beta$ enhancement when we increase the value of $\tan\beta$, the contribution due to heavy Higgs increases but simultaneously the vlaue of $m_{\mu}^{LE}$ decreases. Therefore the modification of muon yukawa coupling and the gauge coupling remains under control.\\

\begin{figure}[h!]
	\centering
\subfloat[]	{\includegraphics[width=0.35\textwidth]{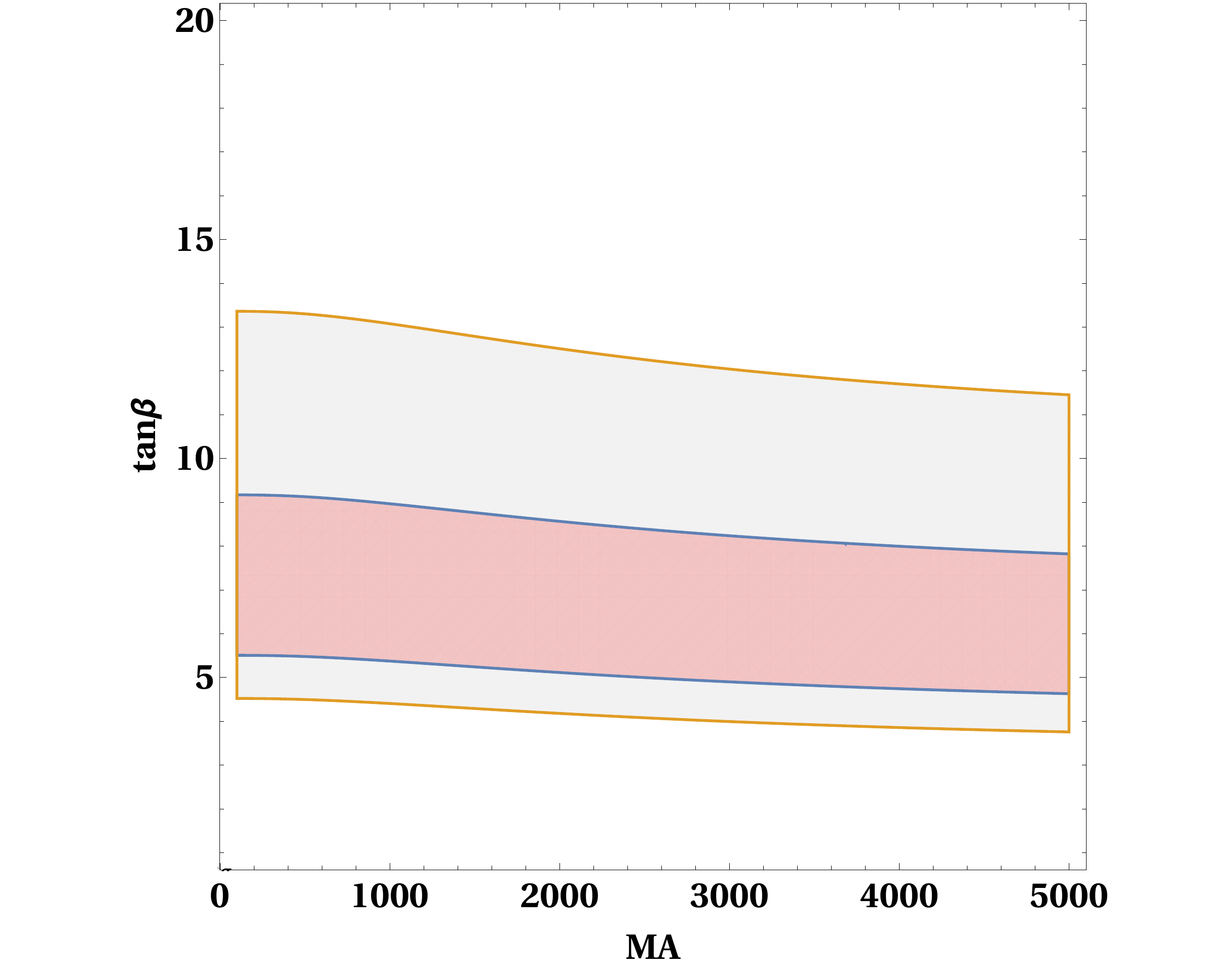}} 
\subfloat[]	{\includegraphics[width=0.35\textwidth]{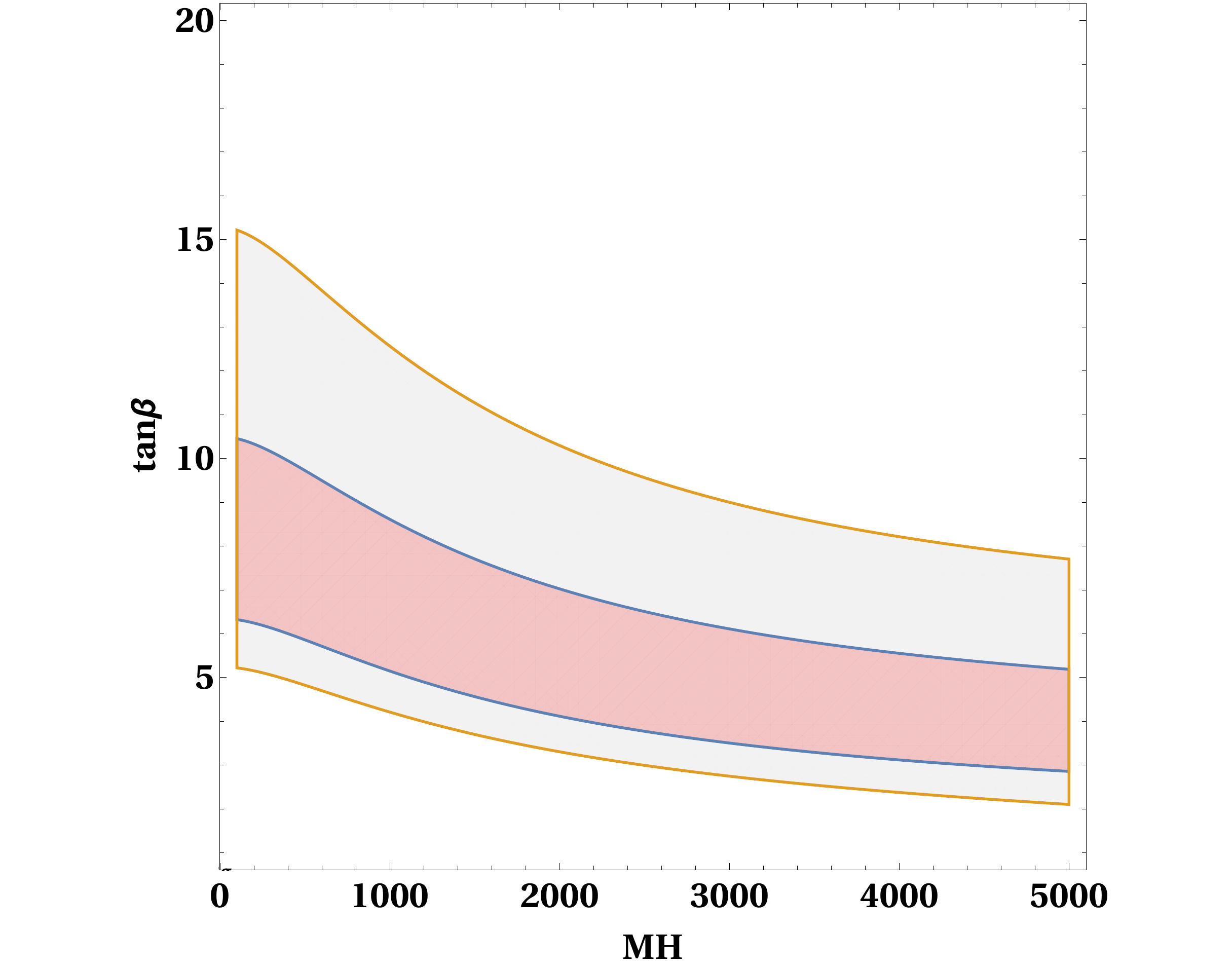}}
\subfloat[]	{\includegraphics[width=0.35\textwidth]{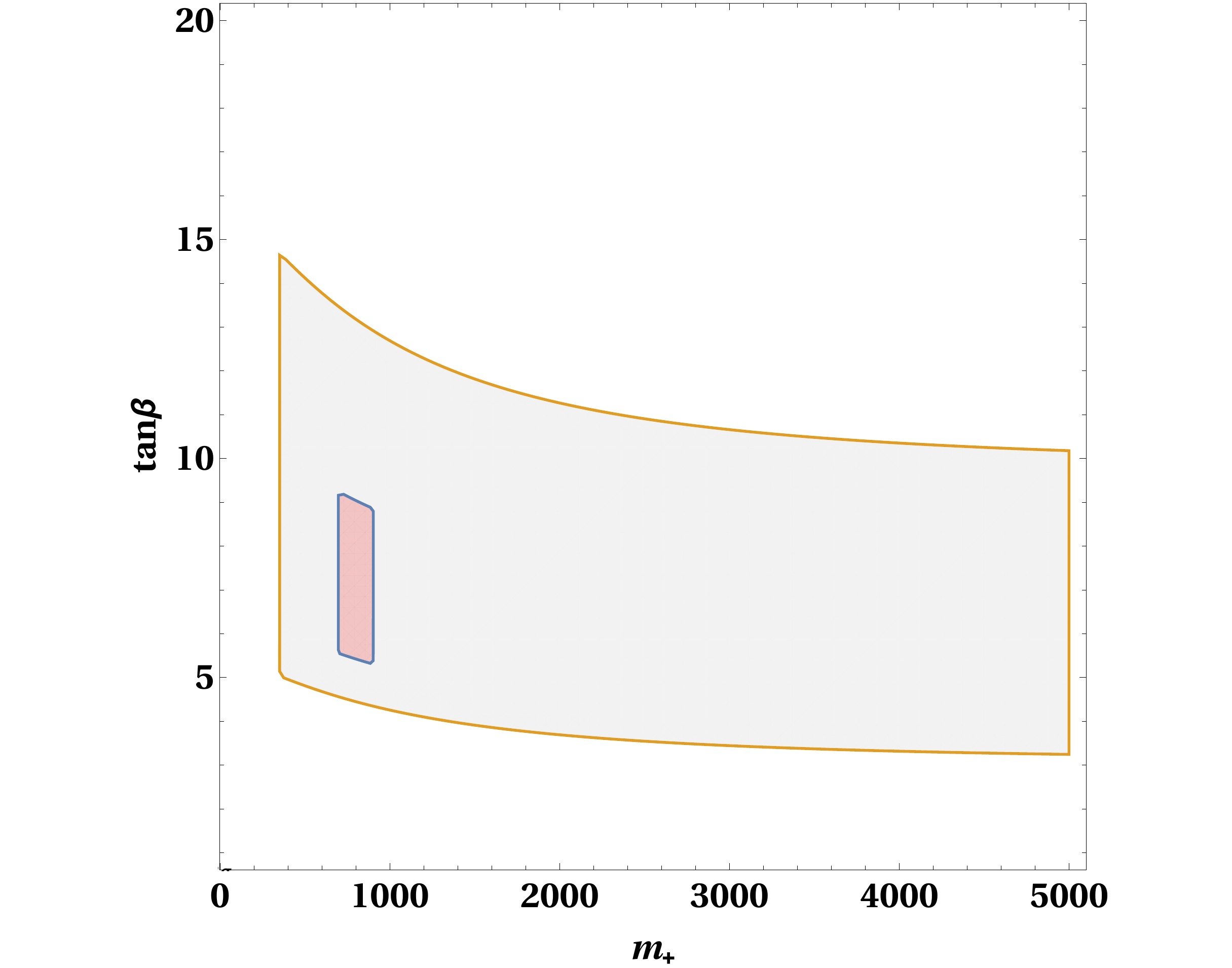}}
	\caption{{\emph{The allowed parameter space in heavy scalar - $\tan\beta$ plane for reproducing the correct value for the muon anomalous magnetic moment in $\mu$2HDM+VLLs. We show the constraints imposed by agreement with the $(g-2)_{\mu}$ at 1$\sigma$ (light red) and 2$\sigma$ (gray).} }}
	\label{fig:matanbeta}
\end{figure}

In figure\eqref{fig:matanbeta} we have plotted the allowed $1\sigma$ and $2\sigma$ region of parameters space satisfying $\Delta^{exp}_{\mu}$  in the heavy Higgs mass - $\tan\beta$ plane. It is  needed to mention here that the limit vary significantly with the assumed pattern of branching ratios of new leptons to $W$, $Z$ and $h$ \cite{Dermisek:2014qca}. \\

\begin{figure}[h!]
	\centering
\subfloat[]	{\includegraphics[width=0.35\textwidth]{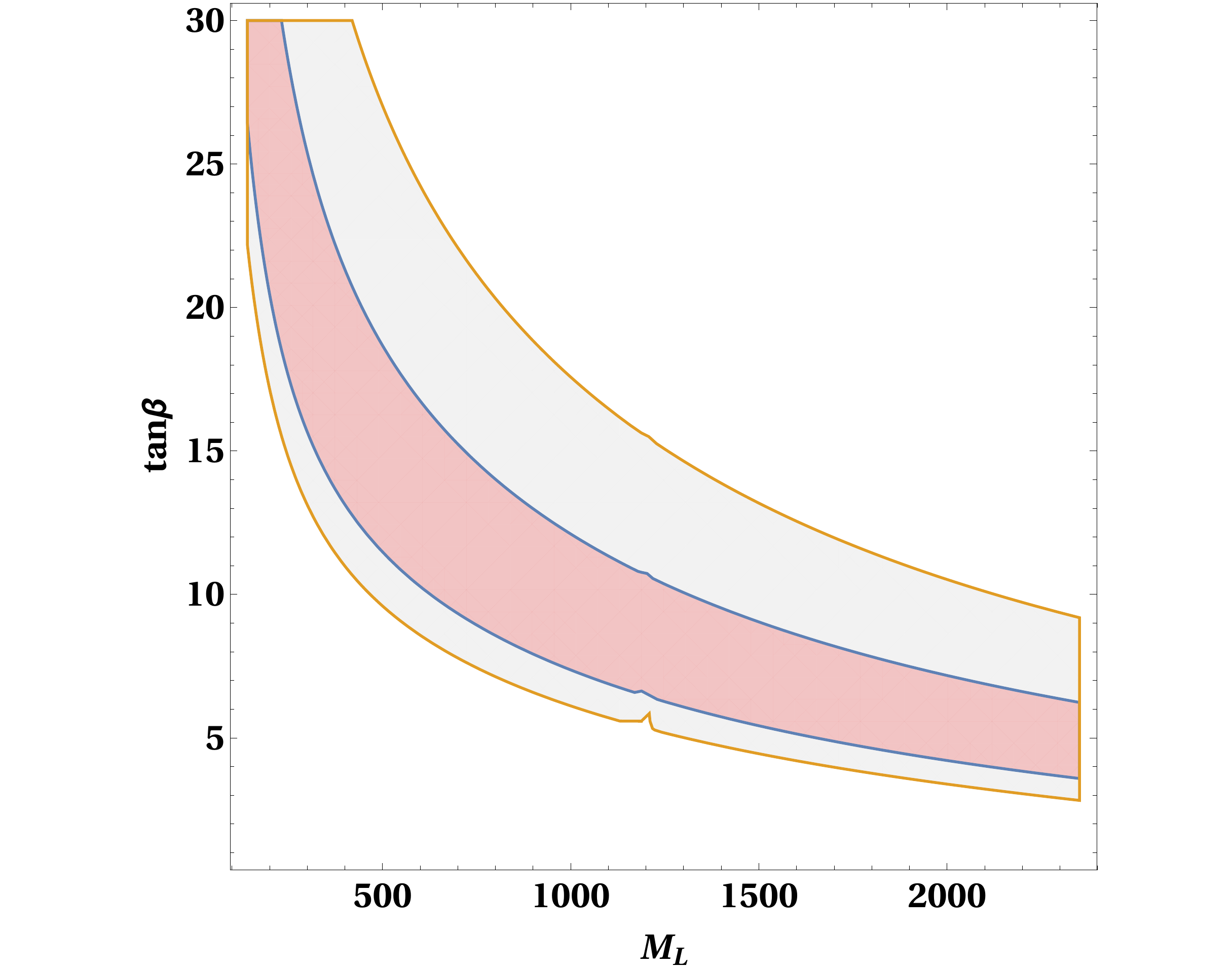}\label{fig:A}}
\subfloat[]	{\includegraphics[width=0.35\textwidth]{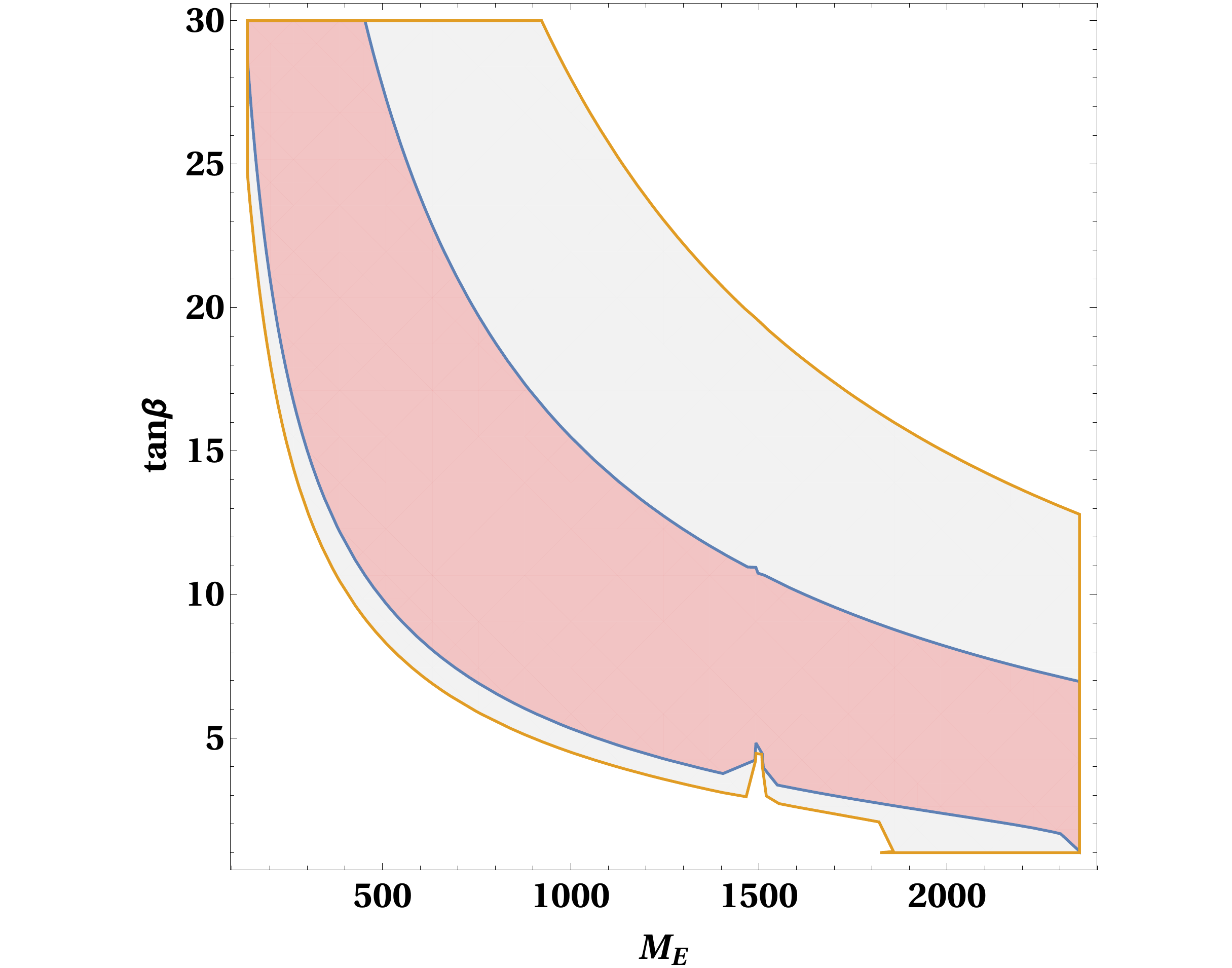}\label{fig:B}}
\subfloat[]	{\includegraphics[width=0.35\textwidth]{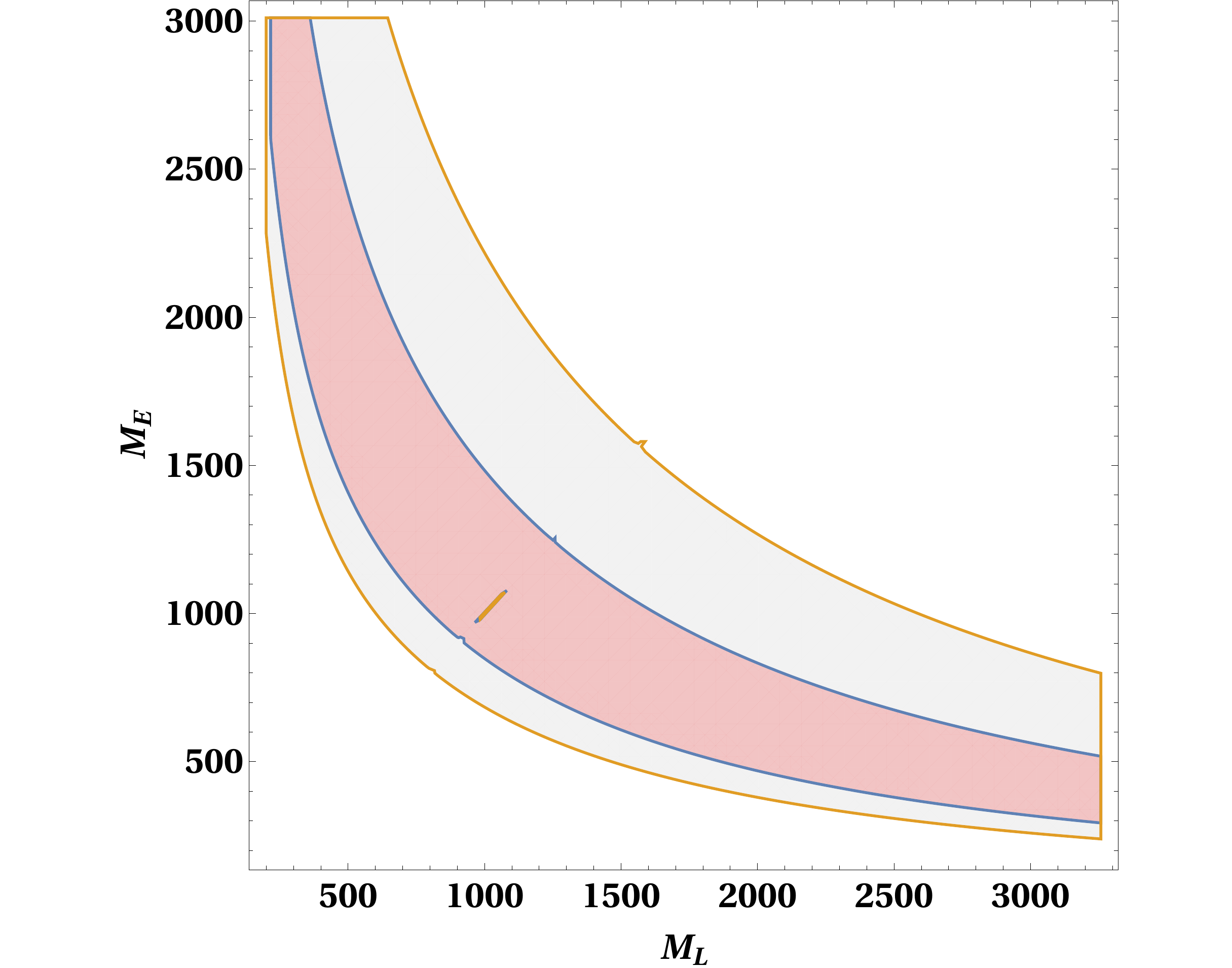}\label{fig:C}}

\caption{{\emph{The allowed parameter space in Vector like leptons for reproducing the correct value for the muon anomalous magnetic moment in $\mu$2HDM+VLLs. We show the constraints imposed by agreement with the $(g-2)_{\mu}$ at 1$\sigma$ (light red) and 2$\sigma$ (gray).} }}
	\label{fig:tanbmlme} 
\end{figure}

Taking forward our analysis  we have created the parameter spaces as a result of our investigation shown in Fig.\eqref{fig:tanbmlme}  where we have illustrated the acceptable region after the  restriction of $(g-2)_{\mu}$ for the $\mu$2HDM + VLLs architecture in several parametric planes. The colour conventions are same as in the previous figure, with the grey region representing the 2$\sigma$ permissible region and the bright red coloured region representing the 1$\sigma$ allowed region of the following parameter space. We can see from Fig.\eqref{fig:B}  that the higher values of $\tan\beta$, are not allowed. This is because of the fact that after that certain value of $\tan\beta$ the decreasing effect of $m_{\mu}^{LE}$ is so high that it can annihilate the enhancement effect due to $\tan\beta$ and decrease the overall contribution coming from heavy Higgs. The balancing effect of $\tan^2\beta$ and $m^{LE}_{\mu}$ generate specific parameters space for $(g-2)_{\mu}$.One more thing we have to mention here is that in our model we successfully generate the parameter space in low $\tan{\beta}$ region which is obviously much better than the work done by \cite{Abe:2017jqo}. Also in the previous work \cite{Abe:2017jqo} they had only shown the allowed parameter space of $M_H$ with $\tan_\beta$, where as we can show all other allowed parameter space relevant to our study. If we look at  the Fig. \eqref{fig:A} and in \eqref{fig:B} it is showing the anti-correlation between the $\tan\beta$ and ML (ME), for higher value of ML(ME) we need lower values of $\tan\beta$. In Fig. \eqref{fig:C} the correlation between the VLLs mass gives us an idea that to satisfy the $(g-2)_\mu$ anomalies, we can take only a certain combination of VLLs masses. At last, our findings in Fig.\eqref{fig:tanbmlme} highlight the characteristics of the model parameters that account for the $(g-2)_{\mu}$ anomaly.

%%%%%%%%%%%%%%%%%%%%%%%%%%%%%%%%%%%%%%%%%%%%%%%%%%%%%%%%%%%%%%%%%%%%%%%%%%%%%%%%%%%%%%%%%%%%%%%%%%%%%%%%%%%%%%%%%%%%%%%%%%%%%%%%%%%%%%%%%%%%%%%%%%%%%%%%%%%%%%%%%%%%%%%%%%%%%%%%%%%%%%%%%%%%%%%%

%%%%%%%%%%%%%%%%%%%%%%%%%%%%%%%%%%%%%%%%%%%%%%%%%%%%%%%%%%%%%%%%%%%%%%%%%%%%%%%%%%%%%%%%%%%%%%%%%%%%%%%%%%%%%%%%%%%%%%%%%%%%%%%%%%%%%%%%%%%%%%%%%%%%%%%%%%%%%%%%%%%%%%%%%%%%%%%%%%%%%%%%%%%%%%%%
\section{Collider Study} \label{collider}
	
In this section we shall discuss the detector level simulation corresponding to VLLs (vector like leptons) in the multi-lepton channels. We have compared the outcomes of the cut based analysis and also multivariate analysis methods while analysing this model. There are many phenomenological studies \cite{Bissmann:2020lge,Bhattiprolu:2019vdu,Kumar:2015tna} on VLLs in the literature, but for the simulation purpose we took the inspiration from the CMS study \cite{CMS:2019hsm} for the search of VLLs. For the simulation purpose we have generate both signal and background events and compute the cross section in {\tt MG5 aMC@NLO}\cite{Alwall:2014hca} at the leading order (LO). The default PDF set NNPDF2.3LO\cite{Ball:2012cx} has been used for the event generation and computing the cross section. We have consider the main SM background for the multi-lepton chanels are $VVV$, $VV$, $t\bar{t}V$, $t\bar{t}h$  and $hV$ (where $V= W,Z$ boson). For each background and as well as signal processes we generate at least $10^5$ events. The showering and hadronisation of the produced events are then processed within the {\tt Pythia 8.2} \cite{Sjostrand:2014zea}. Then for the purpose of detector level simulation we use {\tt Delphes 3.4}\cite{deFavereau:2013fsa}, a fast detector simulation package. We use CMS card to reconstruct jets, electrons, muons, missing energy within the {\tt Delphes 3.4}. We use the anti-kt algorithm for clustering the jets with a radius parameter $R= 0.4$ applying the {\tt FastJet} package \cite{Cacciari:2011ma}. We have shown our simulation analysis for the centre of mass energy of 13 TeV and integrated luminosity of 139 $fb^{-1}$.\\

In our analysis we consider the lighter leptons i.e. $e$ and $\mu$ to get the multi-lepton signal and put the object cuts on them. These objects are reconstructed with the identification efficiency of default CMS card. The events with multi-lepton signal then sorted in two different final state comprising of either two leptons or three leptons. The leading lepton must have to pass the trigger criteria mentioned in ref. \cite{CMS:2019hsm} to qualify as selected event. The restriction on missing energy, $\cancel{E_T}$ ($\cancel{E_T} > 150$ GeV) gives us a better discrimination of signal than SM background. The events with two opposite sign same flavour leptons are labeled as "OS". Here we imposed a condition on invariant mass of the leptons, $M_{ll}$ for the signal region comprising of two opposite sign leptons. We have discarded the pair with $M_{ll}$ within 15 GeV of $M_Z$ for the cut based analysis. This reduces the SM background events which contains leptons coming from $Z$ boson. So the two distinct signal regions are, $3l$ and $2l$-$OS$.

After all these cuts implemented on each of the above mentioned signal region, a minimum bound on $L_T$ has been set to get a significant excess of signal events over the SM background. $L_T$ is defined as the sum of the transverse momentum of all the signal leptons.
\begin{eqnarray}
	L_T = \sum_{l=e,\mu} p_{T}(l) . \nonumber
\end{eqnarray}

We have chosen different minimum cuts on $L_T$ for different signal regions to get clear signal. This optimization of $L_T$ is given based on our simulation. We note in passing that this Monte Carlo simulations are only an approximation of the actual experimental capability. In general, our findings shows that for a higher value of VLLs mass, at edge of the reach, the lower bound on $L_T$ can be increased to optimize the signal. 

We use transverse mass ($m_T$) as a distinguishing variable in multivariate analysis along with all of these variables mentioned above. This variable is only used in the three-lepton final state because it is a good discriminator in that SR. The variable ($m_T$) is defined as $m_T = \sqrt{2 p_T^{miss} p_T^l [1 - \cos(\Delta\Phi_{m_T})]}$, where $p_T^l$ refers to the $p_T$ of the lepton that is not part of the OS pair closest to the Z boson mass and $\Delta\Phi_{m_T}$ is the difference in azimuth angle between $\vec{p}_T^{~miss}$ and $\vec{p}_T^{~l}$.

\subsection{Cut Based Analysis}

\begin{figure}[!h]
	\centering
	\includegraphics[scale=0.85]{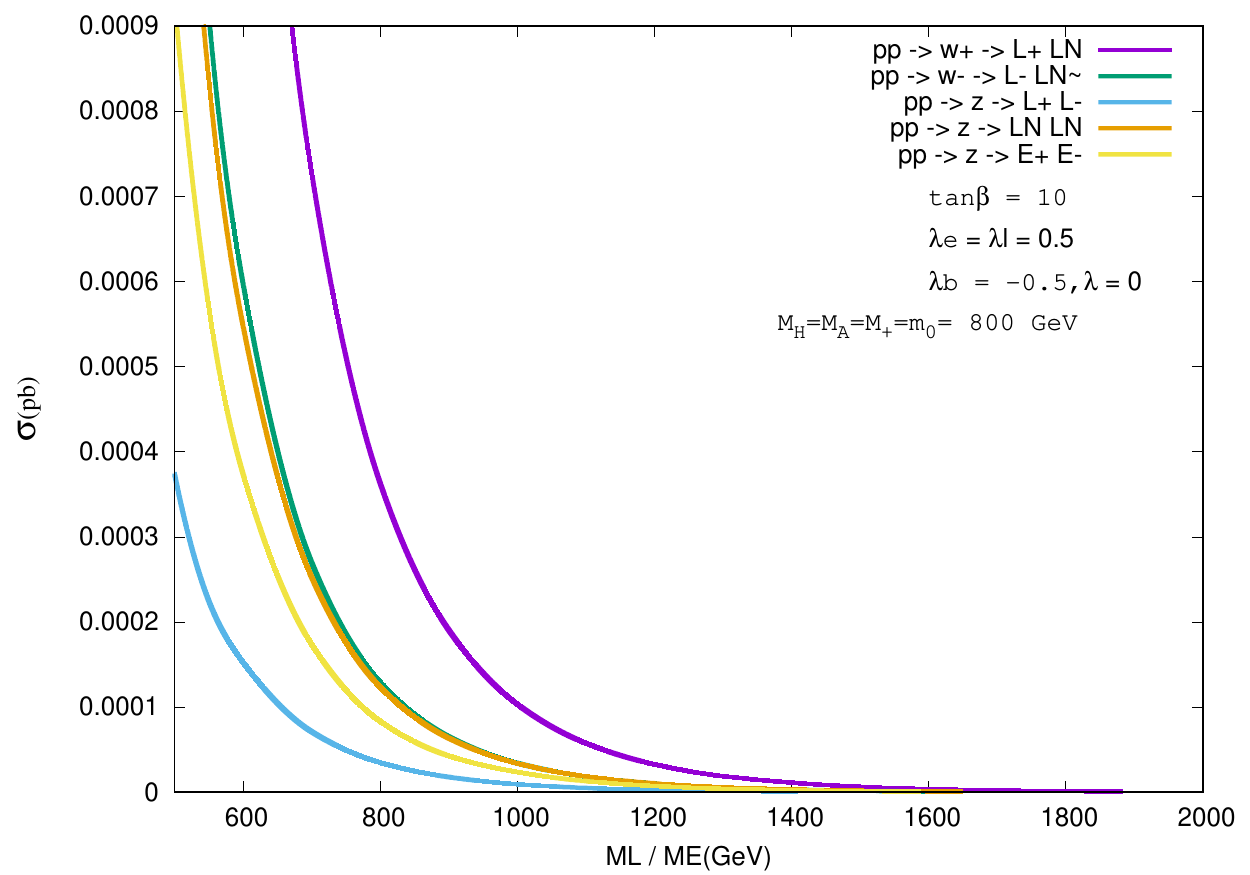}
	\caption{The production cross section of different VLL. We have mentioned the different parameter values which have been considered in order to produce the vector like leptons. }
	\label{fig1}
\end{figure}

In the fig.\eqref{fig1} we have shown the change in production cross section of different VLL with respect to their masses. From the fig.\eqref{fig1} we can infer that the production cross section of $L^+ LN$ is the greatest among all the VLL production. Despite the fact that process $pp \rightarrow W^- \rightarrow L^- \overline{LN}$ and process $pp \rightarrow W^+ \rightarrow L^+ LN$ is conjugate, the production cross section of $L^- \overline{LN}$ is relatively small. This is due to the fact that the production cross section of $W^-$  is nearly 3 nb less than the production cross section of $W^+$ \cite{ATLAS:2016fij}.  Here $L^-$ and $ LN$ are the particle from the VLL doublet and $E^-$ is the particle from the singlet. For this reason further detector level simulations have been done only taking this process into account. $L^+$ dominantly decays into $\mu^+ H$, $\mu^+ A$, $H^+ \nu_{\mu}$ and $E^+ \gamma$ channels, where $H$, $A$, $H^+$ are the heavy CP even, CP odd and heavy charged Higgs respectively. Likewise, the particle $LN$ decays mainly into $\mu^- H^+$, $W^+ \mu^- H$ etc. channels.
The heavy neutral CP even Higgs decays to SM model like Higgs and other channels containing jets. Charged Higgs decays dominantly in the final state comprising of jets. This decay topology gives us a dominat two leptons final state with a lesser amount of final state comprising of three leptons.\\

\FloatBarrier

\begin{table}[h!]	
	\resizebox{\textwidth}{!}{
		\begin{tabular}{||c|c|c|c|c|c|c|c|c|c|c|c||}
			\hline \hline
			BPs & $\tan\beta$ & \makecell{$M_H$ \\ (GeV)} & \makecell{$M_A$ \\ (GeV)} & \makecell{$M_{H^\pm}$ \\ (GeV)} & $\beta_5$ &\makecell{$M_L$ \\ (GeV)} & \makecell{$M_E$ \\ (GeV)} & $\lambda$ & $\lambda_e$ & $\lambda_l$ & $\lambda_b$\\	
			\hline 
			BP-I & 10 & 301 & 321 & 320 & 3.02 & 650 & 600 & 0.0 & 0.41 & 0.110 & -0.59 \\ 
			\hline
			BP-II & 10 & 467 & 498 & 499 & 2.18 & 1000 & 900 & 0.0 & 0.35 & 0.440 & -0.46 \\ 
			\hline
			BP-III & 30 & 318 & 326 & 324 & 5.07 & 800 & 750 & 0.0 & 0.44 & 0.422 & -0.56 \\ 
			\hline
			BP-IV & 30 & 424 & 454 & 495 & 3.24 & 1100 & 1000 & 0.0 & 0.48 & 0.760 & -0.58 \\
			\hline \hline 
	\end{tabular} } 
	\caption{The parameter values we have considered in order to construct the BPs.} 
	\label{table1}
\end{table}

\FloatBarrier

We now introduce some benchmark points (BPs) to analyse the results further. For the purpose of illustration BPs are taken for the two different $\tan\beta$ values and for each $\tan\beta$ we have taken two different mass of VLL. All other parameters are shown in table\eqref{table1}. All BPs are chosen such that, they satisfy all constraints coming from $S,T,U$ parameter, $b \rightarrow s \gamma$ decay etc. which are mentioned in the previous section\eqref{constraints}.\\

\begin{figure}
	\centering
	\begin{minipage}{.5\textwidth}
		\centering
		\includegraphics[width=1.\linewidth]{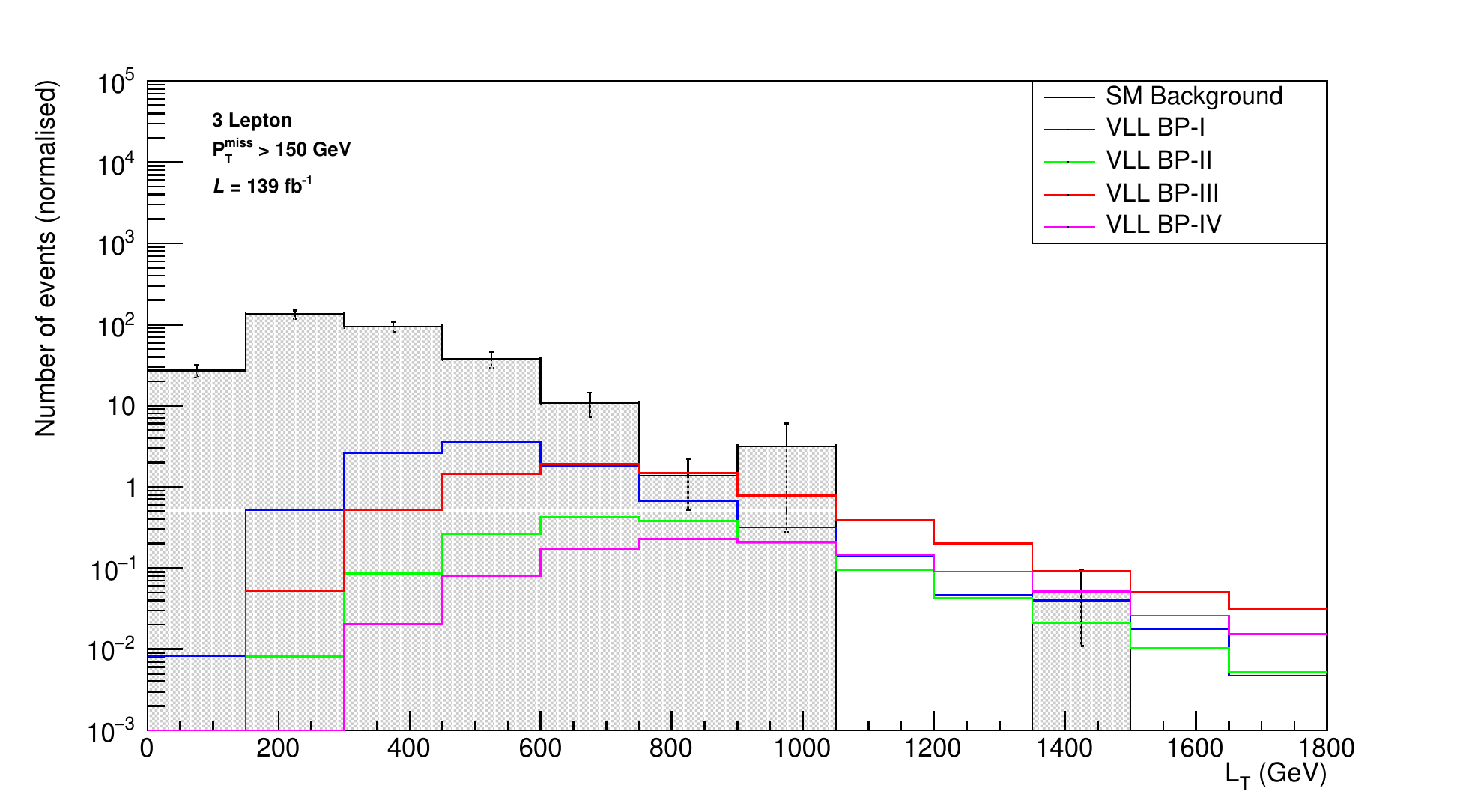}
		\caption{Final state comprising of three leptons}
		\label{fig2}
	\end{minipage}%
	\begin{minipage}{.5\textwidth}
		\centering
		\includegraphics[width=1.\linewidth]{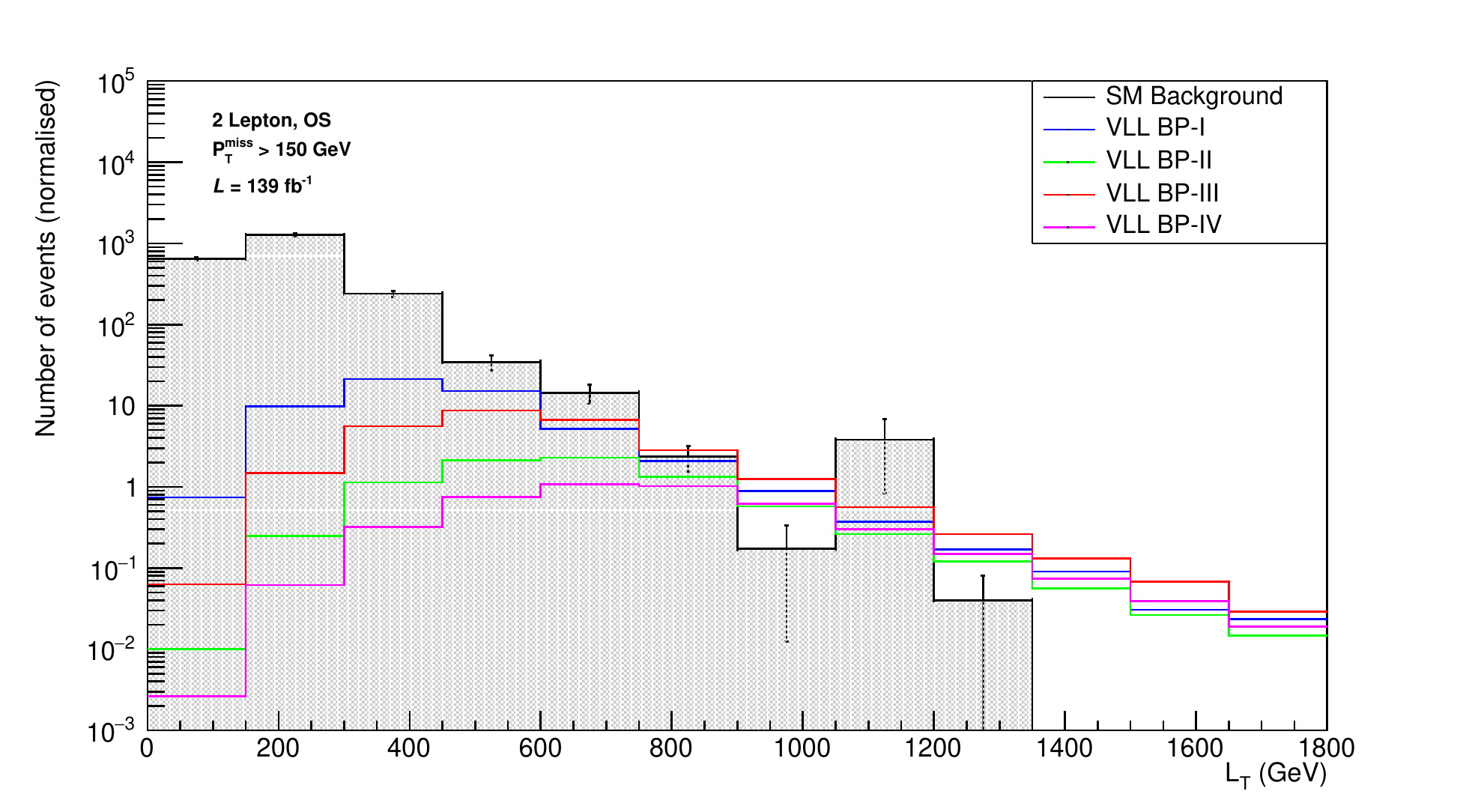}
		\caption{Final state comprising of two opposite sign leptons}
		\label{fig3}
	\end{minipage}
	\caption{We have displayed $L_T$ distribution in this plot for the final state comprising of three and two opposite sign leptons. The shaded region represent the SM background for this final state. The distribution of $L_T$ for the BPs are shown in different colors. }
	%The binning uncertainties has been shown with error bar for the SM background.}
	\label{figcut}
\end{figure}

In fig.\eqref{fig2} we have shown the distribution of $L_T$ for the final state comprising of 3 leptons. From this distribution we can infer that we have to put a stronger minimum cut on $L_T$ ($L_T >1000$ GeV)	to get strong VLL signal event over background. We have shown the normalised event number after passing all the cuts in the table\eqref{table2} for the 3 lepton final state.\\

\FloatBarrier

\begin{table}[h!]
	\begin{center}

		\begin{tabular}{|c|cc|cc|}
			\hline
			\multirow{2}{*}{BPs}                         & \multicolumn{2}{c|}{Three leptons (3l)}         & \multicolumn{2}{c|}{Two opposite sign leptons (2l-OS)} \\ \cline{2-5} 
			& \multicolumn{1}{c|}{$N_{Sig}$} & Significance & \multicolumn{1}{c|}{$N_{Sig}$}    & Significance   \\ \hline
			BP1                                           & \multicolumn{1}{c|}{0.329}     &  \multicolumn{1}{c|}{0.54}       & \multicolumn{1}{c|}{1.589}        &  \multicolumn{1}{c|}{0.94}       \\ \hline
			BP2                                           & \multicolumn{1}{c|}{0.234}     &  \multicolumn{1}{c|}{0.45}            & \multicolumn{1}{c|}{1.069}        &  
			\multicolumn{1}{c|}{0.70}         \\ \hline
			BP3                                           & \multicolumn{1}{c|}{1.002}     & \multicolumn{1}{c|}{0.98}             & \multicolumn{1}{c|}{2.331}        &  
			\multicolumn{1}{c|}{1.23}         \\ \hline
			BP4                                           & \multicolumn{1}{c|}{0.409}     & \multicolumn{1}{c|}{0.61}             & \multicolumn{1}{c|}{1.225}        &  
			\multicolumn{1}{c|}{0.78}         \\ \hline
			\begin{tabular}[c]{@{}c@{}}SM ($N_{SM}$)\\ background \end{tabular} & \multicolumn{1}{c|}{0.04}      &  
			\multicolumn{1}{c|}{-}            & \multicolumn{1}{c|}{1.258}        & \multicolumn{1}{c|}{-}               \\ 
			\hline
		\end{tabular} 
	\end{center}
	\caption{In this table we have displayed the normalised event number ($N_{Sig}$)  and the significance $\left( \frac{N_{Sig}}{\sqrt({N_{Sig}+N_{SM})}}\right) $at 139 $fb^{-1}$ for the signal (BPs) and background event for the final state comprising of 3 leptons and two opposite sign leptons.}
	\label{table2}
\end{table}			
\FloatBarrier	

Next we have discussed the signature of final state comprising of 2 leptons. In fig.\eqref{fig3} we have displayed the normalised event distribution with the final state comprising of 2 opposite sign leptons. We have put a cut on $M_{ll}$ to exclude the event that can emerge from $Z$ boson. In this signal region we have put a minimum bound on $L_T$ ($L_T > 900$ GeV) to obtain the signal event. The normalised event numbers are shown in table\eqref{table2}. From this table\eqref{table2} and the plots [fig.\eqref{fig3},fig.\eqref{fig2}] we can say that for the low $\tan\beta$(~10) region, 2l-OS is the favorable signal channel for VLLs signature, where as at high $\tan\beta$(~30) region both the channel 3l and 2l-OS  are almost equally favorable (as far the significance concern). But the problem is, we can not discriminant the signal from the SM background in a very precise manner. Additionally, the signal strength is also weaker in the CBA due to the small production cross section of the VLL (see Fig.\eqref{fig1}). That is why, we must move to MVA from CBA in order to accomplish all the above mentioned drawbacks. \\ 		

\subsection{Multivariate Analysis}

We present a multivariate analysis in this subsection for better signal to background differentiation, which leads to an increment in significance. In the {\tt TMVA} \cite{Voss:2007jxm} framework within the {\tt ROOT} \cite{Antcheva:2009zz}, we implement the Boosted Decision Tree (BDT) algorithm for the Multivariate Analysis (MVA). To achieve substantial significance in the cut-based analysis discussed in the preceding subsection, we have identified an cut value for the variables. The MVA technique is a powerful tool for obtaining the optimal sensitivity for a given set of parameters for this purpose. We use four variables for each signal region for the MVA. In the table\eqref{imp} we have shown those variables along with the importance of those variables in the BDT response. These variables have been determined by comparing the background distributions with the signal trained for $M_L= 1007$ GeV with $\tan\beta$= 10 and $M_E = 850$ GeV at $\sqrt{s}=13$ TeV. We have incorporated a new variable, transverse mass ($m_T$), as the discriminating input variable for the BDT for the final state comprised of three leptons, in addition to the variables presented in the cut based analysis. 

\begin{figure}
	\centering
 \begin{minipage}{.9\textwidth}
		\centering
		\includegraphics[width=.9\linewidth]{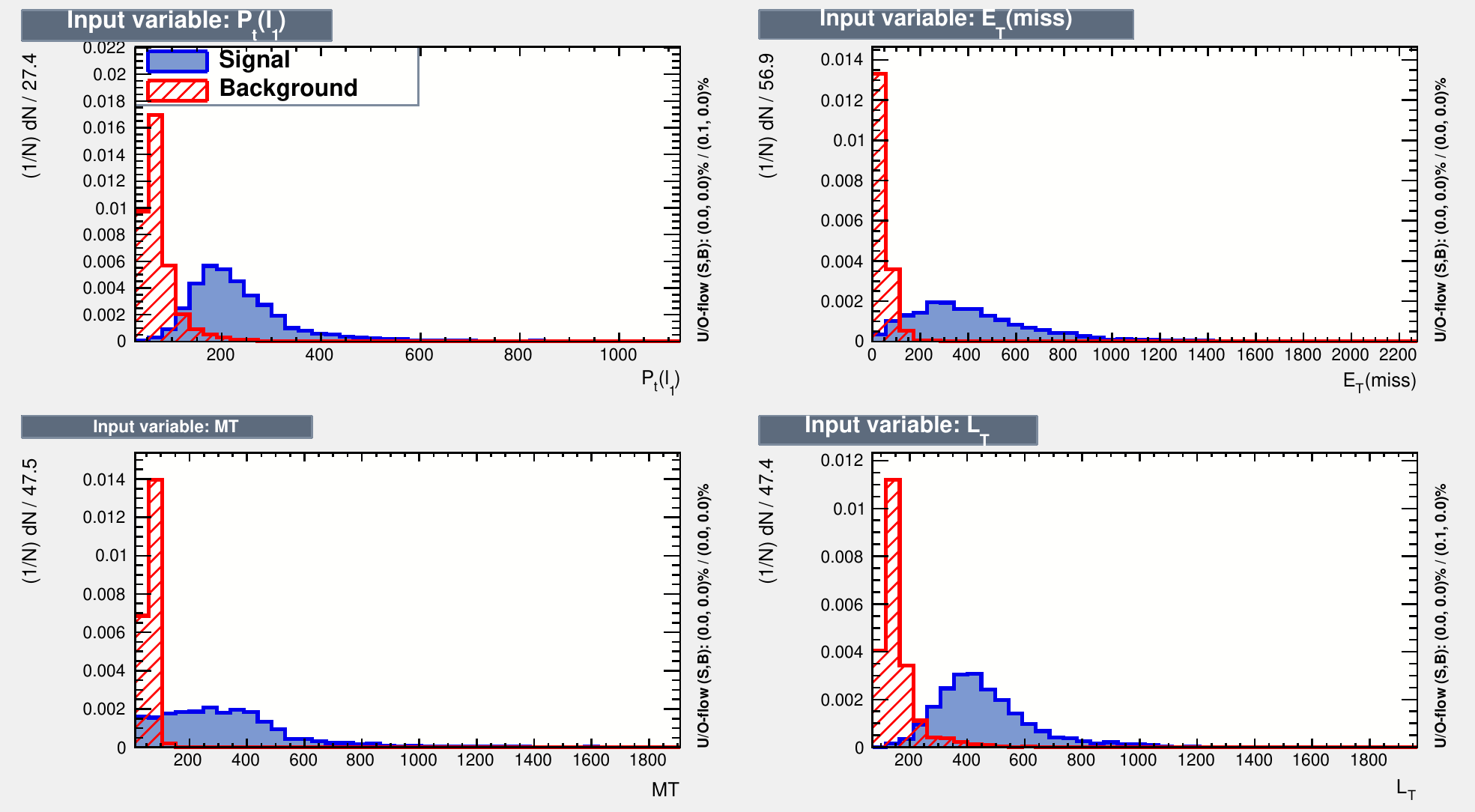}
		\caption{Final state comprising of three leptons}
		\label{fig5}
\end{minipage}
	
	\begin{minipage}{.9\textwidth}
		\centering
		\includegraphics[width=.9\linewidth]{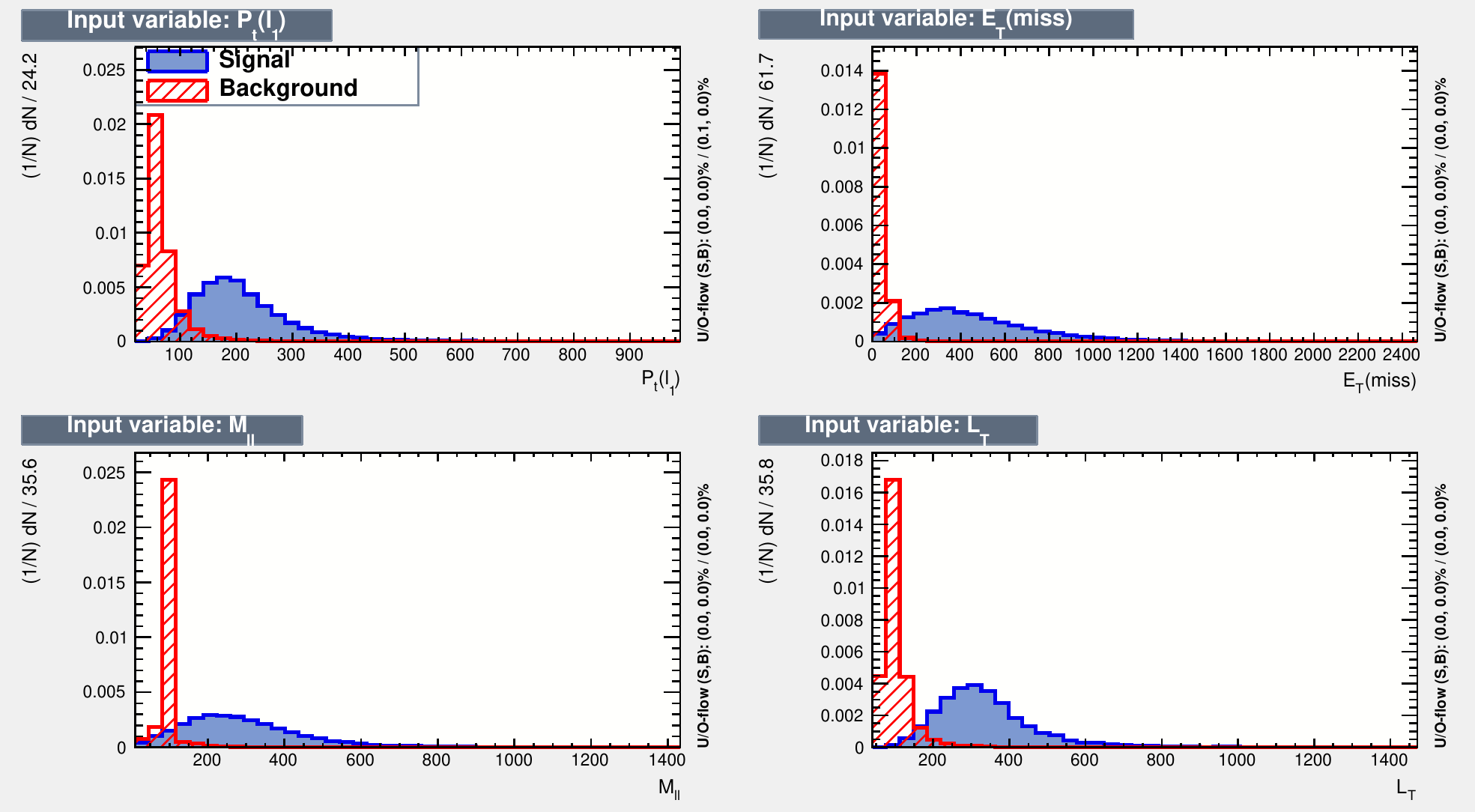}
		\caption{Final state comprising of two opposite sign leptons}
		\label{fig6}
	\end{minipage}
	\caption{The signal(blue) and background(red) distributions of the input variables used for MVA have been shown in this figure. For the final state comprising of three leptons we have used the transverse mass ($m_T$) as one of the input discriminating variable. The invariant mass of two final lepton is only used as discriminating variable for two lepton final state. Other variables, which are  described in the earlier section are same for these two SRs.}
	\label{figSvsB}
\end{figure}

As seen in fig.\eqref{figSvsB}, where we have presented the normalised signal and background event distributions, each of these variables has a respectable level of discriminating power. It's worth noting that the four variables utilised here may not be the best, and there's always the possibility of improving the analysis by making better variable choices.
We utilised these simple kinematic variables in our study since they are less correlated and have significant discriminating power. To better the analysis, a more focused MVA can be incorporated with distinct sets of variables for different parameter points. \\

 \begin{table}[h!]
	\begin{center}
		\begin{tabular}{||c|c|c|c||}
			\hline
			\multicolumn{2}{|c|}{Three leptons (3l)}                    & \multicolumn{2}{c|}{Two opposite sign leptons (2l-OS)}        \\ \hline
			Variable & Importance & Variable & Importance      \\ 
			\hline
			$P_{t}(l_1)$ & $2.289 \times 10^{-1}$ & $P_{t}(l_1)$ & $2.194\times 10^{-1}$ \\ \hline
			$E_T^{miss}$ & $2.803\times 10^{-1}$  & $E_T^{miss}$ & $2.539\times 10^{-1}$ \\ \hline
			$m_T$    & $2.306\times 10^{-1}$ & $M_{ll}$ & $2.753\times 10^{-1}$ \\ \hline
			$L_T$    & $2.602\times 10^{-1}$ & $L_T$    & $2.514\times 10^{-1}$ \\ \hline
		\end{tabular} 
	\end{center}
	\caption{The relative importance of the input variables utilised in MVA with $M_L= 1007$ GeV with $\tan\beta$= 10 and $M_E = 850$ GeV at $\sqrt{s}=13$ TeV. This could be different for different sets of parameters. }
	\label{imp}
	
\end{table}

\begin{figure}
	\centering
	\begin{minipage}{.5\textwidth}
		\centering
		\includegraphics[width=.99\linewidth]{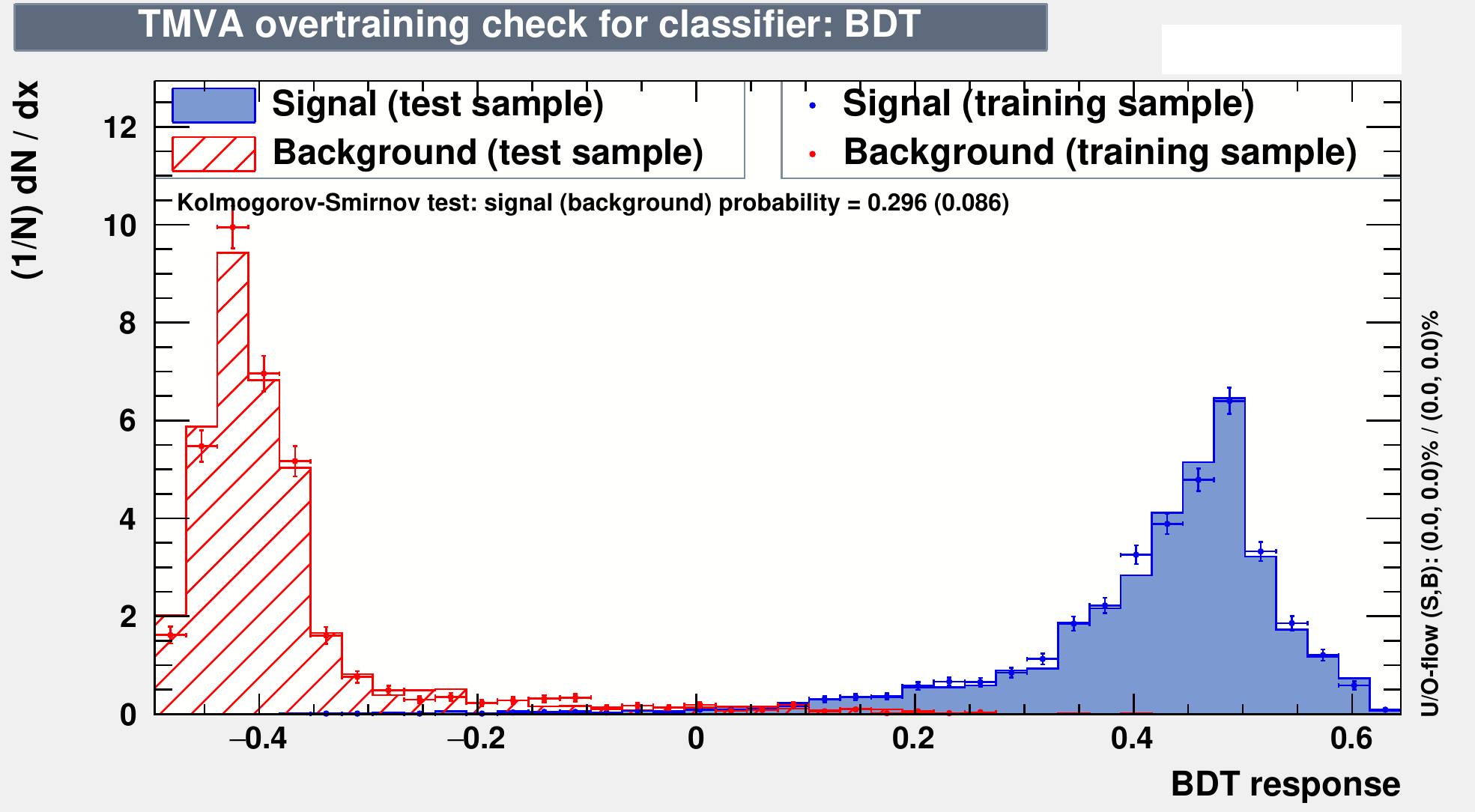}
		\caption{Final state comprising of three leptons}
		\label{fig7}
	\end{minipage}%
	\begin{minipage}{.5\textwidth}
		\centering
		\includegraphics[width=.99\linewidth]{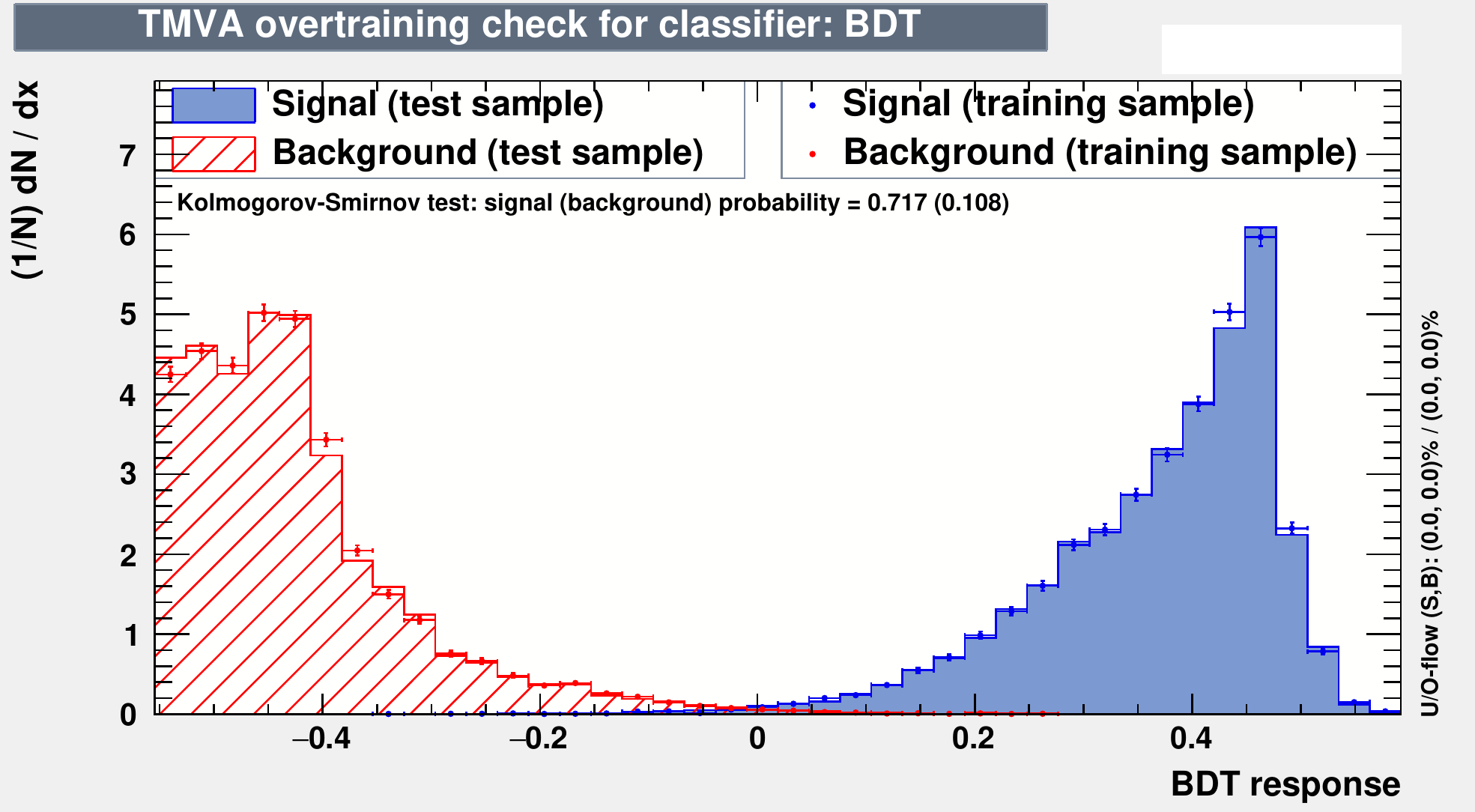}
		\caption{Final state comprising of two opposite sign leptons}
		\label{fig8}
	\end{minipage}
	\caption{Overtraining check of the BinfarDT response for the parameter set $M_L= 1007$ GeV, $\tan\beta$= 10, $M_E = 850$ GeV at $\sqrt{s}=13$ TeV.}
	\label{figovertrain}
\end{figure}

The Kolmogorov-Smirnov (KS) test can be used to determine whether or not a test sample is over-trained. In general, if the KS probability is somewhere between 0.1 and 0.9, the test sample is not over-trained. A critical KS probability value greater than 0.01 assures that the samples are not over-trained in most scenarios. The KS probability values for the signal and background of the BDT response are presented in fig.\eqref{figovertrain} , indicating that neither the signal nor the background samples have been over-trained. We have made sure that we do not get over-trained on any of the parameter points we have mentioned. As seen in fig.\eqref{figovertrain}, the signal and background samples in this BDT output are well separated, allowing us to considerably enhance the signal significance by applying an appropriate BDT cut.

\FloatBarrier
\begin{table}[h!]
	\begin{center}

		\begin{tabular}{||c|c|c|c|c|c|c|c|c||}
			\hline
			\multirow{2}{*}{} & 
			\multicolumn{4}{|c|}{Three leptons (3l)}                    & \multicolumn{4}{|c|}{Two opposite sign leptons (2l-OS)} \\ 
			\cline{2-9}
			\makecell[c]{$M_L$ \\(GeV)} & \makecell[c]{BDT \\cut value} & $N_S$ & $N_B$ & \makecell[c]{Significance \\$\frac{N_{S}}{\sqrt({N_{S}+N_{B})}}$} & \makecell[c]{BDT \\cut value} & $N_S$ & $N_B$ & \makecell[c]{Significance \\$\frac{N_{S}}{\sqrt({N_{S}+N_{B})}}$} \\ 
			\hline
			1007 & 0.0810 & 975.281 & 11.843 & 31.04 & -0.0262 & 994.942 & 6.750 & 31.44 \\
			\hline
			1207 & -0.0339 & 998.113 & 5.400 & 31.51 & -0.0086 & 999.167 & 1.687 & 31.58 \\  
			\hline
			1409 & -0.0999 & 998.765 & 0.543 & 31.59 & -0.0818 & 999.462 & 0.366 & 31.61 \\
			\hline
			1606 & -0.2435 & 1000.000 & 1.077 & 31.61 & -0.0918 & 999.705 & 0.427 & 31.61\\ 
			\hline		
				\hline
			\multirow{2}{*}{} & 
			\multicolumn{4}{|c|}{Three leptons (3l)}                    & \multicolumn{4}{|c|}{Two opposite sign leptons (2l-OS)} \\ 
			\cline{2-9}
			\makecell[c]{$\tan\beta$} & \makecell[c]{BDT \\cut value} & $N_S$ & $N_B$ & \makecell[c]{Significance \\$\frac{N_{S}}{\sqrt({N_{S}+N_{B})}}$} & \makecell[c]{BDT \\cut value} & $N_S$ & $N_B$ & \makecell[c]{Significance \\$\frac{N_{S}}{\sqrt({N_{S}+N_{B})}}$} \\ 
					\hline
			08 & -0.0678 & 997.81 & 8.61 & 31.45 & -0.013 & 998.07 & 1.893 & 31.56 \\
			\hline
			12 & 0.1028 & 995.25 & 6.45 & 31.45 & -0.0011 & 998.21 & 1.18 & 31.57 \\  
			\hline
			16 & 0.1009 & 996.61 & 4.86 & 31.49 & 0.0239 & 997.78 & 1.614 & 31.56 \\
			\hline
			20 & -0.0332 & 996.30 & 7.55 & 31.44 & -0.0098 & 998.39 & 0.912 & 31.58\\ 
			\hline		
		\end{tabular} 
	\end{center}
	\caption{Numbers of signal ($N_S$) and background ($N_B$) events for different $M_L$ and $\tan\beta$ after passing the BDT cut have been shown for luminosity $L = 139 fb^{-1}$. }
	\label{vllbdt}
	
\end{table}
\FloatBarrier
We can observe that in the high $\tan\beta$ region, the cut based analysis gives us a better outcome over SM background, whereas even in the low $\tan\beta$ region, BDT gives us a substantial signal significance. The MVA is significantly more effective to analyze the muon specific vector like lepton model relative to cut based analysis. We compute the significance given in table\eqref{vllbdt} for various $M_L$ at $\tan\beta = 10$, $\sqrt{s}$ = 13 TeV and $L = 139 fb^{-1}$, and also shows the significance for different $\tan\beta$ values at a given $M_L = 1200 $GeV with same $\sqrt{s}$ and $L$. From this tables of MVA  we can also conclude that both the 3l and 2l-OS are equally good for our analysis. The MVA significant does not depend on the $M_L$ or $\tan \beta$, this is not an abnormal fact, as we know that MVA maintain a crucial thumb rule that it set the BDT cut in such a way to get that maximum signal significance i.e. to discriminant the signal from the SM background with the significant precision. For that we have to specify some variables with reasonable importance. In this analysis we choose almost same sets of variables for 3l and 2l-OS except the $m_T$ for 3l is replaced by $M_{ll}$ in case of 2l-OS. Using these variables MVA generates BDT cuts in such a way that the signal significance must be the maximum (given in table\eqref{vllbdt}). It's the role of BDT which gives us an idea that both 3l and 2l-OS are equally important to analyse VLLs collider signature.

%%%%%%%%%%%%%%%%%%%%%%%%%%%%%%%%%%%%%%%%%%%%%%%%%%%%%%%%%%%%%%%%%%%%%%%%%%%%%%%%%%%%%%%%%%%%%%%%%%%%%%%%%%%%%%%%%%%%%%%%%%%%%%%%%%

\section{Summary and Conclusion} \label{conclusion}

The constant alignment between the prediction from the standard model (SM) and the experimental data from LHC so far has posed strong challenges for new physics (NP) scenarios beyond the standard model (BSM). But in this smooth path way the measurement of  anomalous magnetic moment of the muon remains one of the continuing deviation from SM expectation\cite{Aoyama:2020ynm}. Stretching this deviation forward the $(g-2)_\mu$ experiment \cite{Fienberg:2019ddu,Sato:2021aor} consolidate the ground for BSM. Without vector-like leptons, the type-I and type-Y models cannot explain the discrepancy and the type II requires light pseudoscalar that are in conflict with BR$(B\rightarrow X_s \gamma)$\cite{Belle:2016ufb,Misiak:2017bgg}. The type-X model can accommodate the discrepancy but requires large values of $\tan\beta$ and also very light pseudoscalar\cite{Chen:2021jok}. The $\mu$2HDM can explain $(g-2)_{\mu}$ but it also requires  very large $\tan\beta$\cite{Abe:2015oca}. After revisiting $\mu$2HDM  in light of the new result reported by the $(g-2)_{\mu}$ collaboration at Fermilab, we have studied the discrepancy of the  $(g-2)_{\mu}$  in the $\mu$2HDM+VLLs with mixing. We restricted our analysis to the alignment limit, where the SM and lightest CP-even Higgs bosons coincide, ensuring agreement with LHC result. Firstly we have analysed the charged VLLs  and extra charged scalar effect on  the higgs diphoton decay. Also we have ensured the effect of modified higgs muon coupling on $h \rightarrow \mu^+ \mu^-$ decay. The parameter space of $\mu$2HDM+VLLs chosen to analyze the $(g-2)_{\mu}$ passed through the constraints from precision electroweak parameters, S and T. Additionally, the Higgs potential's perturbativity, unitarity, and vacuum stability must be respected by keeping the coupling constants within certain bounds. Using these limits we have studied the discrepancy of $(g-2)_{\mu}$ in $\mu$2HDM+VLLs and tried to revel the effect of extra heavy scalar and the vector leptons. Here, we have taken into account the 1-loop contribution from additional scalar and vector leptons. In order to maintain consistency with the perturbative limit, we have taken into account all coupling values up to 1. To explain the discrepancy of $(g-2)_\mu$, the existing $\mu$2HDM  needed a very large value of $\tan\beta$ where as by including the vector lepton with $\mu$2HDM, we can bring down the $\tan\beta $ value to a range between (5-10) for heavy extra scalar and vector leptons masses of the order of (TeV).

In the collider section we have focused our study to the VLL production and its decay to heavy Higgs boson. As $pp \rightarrow L^+ LN$ has the greater production cross section, all the collider study and results are showcased considering this process only. To distinguish the signal from the background, we have used two alternative strategies. Firstly we present a traditional cut-based analysis considering the various kinematical properties of the signal over background. Then, to improve the separation of the signal from the background, we perform a multivariate analysis using a boosted decision tree approach which enables us to obtain a better signal significance and signal strength for both the 3l and the 2l-OS signal final stats.

%%%%%%%%%%%%%%%%%%%%%%%%%%%%%%%%%%%%%%%%%%%%%%%%%%%%%%%%%%%%%%%%%%%%%%%%%%%%%%%%%%%%%%%%%%%%%%%%%%%%%%%%%%%%%%%%%%%%%%%%%%%%%%%%%%%%%%%%%%%%%%%%%%%%
\section*{Acknowledgement}
Abhi Mukherjee would like to thank DST for providing the INSPIRE Fellowship[IF170693]. Jyoti Parasad Saha would like to thanks University of Kalyani for providing personal research grant (PRG) for doing research.

%%%%%%%%%%%%%%%%%%%%%%%%%%%%%%%%%%%%%%%%%%%%%%%%%%%%%%%%%%%%%%%%

\section{Appendix}
\vspace{-5mm}
\subsection{Diagonalizing Mass Matrices}
\vspace{-5mm}
If we consider the limit  $\lambda_E v_1 ,\lambda_E v_1,\lambda_E v_1,\lambda_E v_1 \ll M_L , M_E$ the approximate analytic formulas for diagonalization matrices can be obtained \cite{Dermisek:2013gta,Grimus:2000vj}
\begin{eqnarray}
	U_L =
	\begin{pmatrix}
		1-\frac{v_1^2}{2} \frac{\lambda_E^2}{ M_E^2} & \frac{v_1^2} {2}\Big(\frac{\lambda_E}{M_L} \frac{\bar\lambda M_E + \lambda M_L}{M_E^2 - M_L^2} - \frac{y_{\mu} \lambda_L}{M_L^2}\Big) & v_1\frac{\lambda_E}{M_E} \\
		v_1^2 \frac{\bar\lambda \lambda_E M_L -y_{\mu}\lambda_L M_E}{M_L^2 M_E} &  1-v_1^2  \frac{(\lambda M_E + \lambda M_L)^2}{(M_E^2 - M_L^2)^2}  & v_1  \frac{\bar\lambda M_L + \lambda M_E}{M_E^2 - M_L^2} \\
		-v_1\frac{\lambda_E}{M_E} & -v_1  \frac{\bar\lambda M_L + \lambda M_E}{M_E^2 - M_L^2} & 1-v_1^2 \frac{\lambda_E^2}{ M_E^2} - v_1^2  \frac{(\lambda M_E + \lambda M_L)^2}{(M_E^2 - M_L^2)^2} 
	\end{pmatrix}   \\U_R = 
	\begin{pmatrix}
		1-v_1^2 \frac{\lambda_L^2}{2 M_L^2} & v_1\frac{\lambda_L}{M_L} & v_1^2 \Big(\frac{\lambda_L}{M_E} \frac{\bar\lambda M_L + \lambda M_E}{M_E^2 - M_L^2} + \frac{y_{\mu} \lambda_L}{M_E^2}\Big)   \\
		-v_1\frac{\lambda_L}{M_L} &  1-v_1^2 \frac{\lambda_L^2}{ M_L^2} - v_1^2  \frac{(\lambda M_E + \lambda M_L)^2}{(M_E^2 - M_L^2)^2}   & v_1  \frac{\bar\lambda M_E + \lambda M_L}{M_E^2 - M_L^2} \\
		v_1^2 \frac{\bar\lambda \lambda_L M_E -y_{\mu}\lambda_E M_L}{M_L M_E^2}  & -v_1  \frac{\bar\lambda M_E + \lambda M_L}{M_E^2 - M_L^2} & 1 - v_1^2  \frac{(\bar\lambda M_E + \lambda M_L)^2}{(M_E^2 - M_L^2)^2} 
	\end{pmatrix}
\end{eqnarray}
\subsection{Loop Functions}
The  loop functions for W boson contribution
\begin{align*} 
	F_W(x)&=\frac{4 x^4 - 49 x^3 + 78 x^2 - 43 x +10  + 18 x^3 \ln(x)}{6 (1-x)^4} \\ \nonumber
	G_W(x)&=\frac{-  x^3 + 12 x^2 - 15 x + 4  - 6 x^2 \ln(x)}{ (1-x)^3}
\end{align*}
The  loop functions for Z boson contribution
\begin{align*} 
	F_Z(x)&=\frac{5 x^4 - 14 x^3 + 39 x^2 - 38 x + 8  - 18 x^3 \ln(x)}{12 (1-x)^4} \\ \nonumber 
	G_Z(x)&= - \frac{ x^3 +  3 x - 4  - 6 x \ln(x)}{ 2(1-x)^3}
\end{align*}
The  loop functions for scalar bosons where $\phi=\{ h,H,A \}$
\begin{align*} 
	F_{\phi}(x)&=\frac{x^3 - 6 x^2 + 3 x + 2 + 6 x^3 \ln(x)}{6 (1-x)^4} \\ \nonumber 
	G_{\phi}(x)&=  \frac{ - x^2 +  4 x - 3  - 2 \ln(x)}{ (1-x)^3}
\end{align*}
The  loop functions for $H^{\pm}$ contribution
\begin{align*} 
	F_{H^{\pm}}(x)&=\frac{2x^3 + 3 x^2 - 6 x + 1 - 6 x^2 \ln(x)}{6 (1-x)^4} \\ \nonumber 
	G_{H^{\pm}}(x)&=  \frac{  -x^2 + 1 + 2 x \ln(x)}{ (1-x)^3}
\end{align*}
\vspace{-8mm}

%%%%%%%%%%%%%%%%%%%%%%%%%%%%%% References%%%%%%%%%%%%%%%%%%%%%%%%%%
\vspace{-8mm}
\bibliographystyle{hephys}
\bibliography{reference}

\end{document}